\documentclass[10pt, conference, letterpaper]{IEEEtran}
\IEEEoverridecommandlockouts
\usepackage{cite}

\ifCLASSINFOpdf
\else
\fi
\usepackage{graphicx}
\usepackage{caption}
\usepackage{subcaption}
\usepackage{hyperref}
\usepackage{xcolor}
\usepackage{amsmath}
\usepackage{booktabs}

\usepackage[utf8]{inputenc}
\usepackage[linesnumbered,ruled,vlined]{algorithm2e}

\definecolor{darkgreen}{rgb}{0.0, 0.5, 0.0}

\usepackage[linesnumbered,ruled]{algorithm2e}

\usepackage{amsmath}  
\DeclareMathOperator{\Var}{Var}

\hypersetup{
    colorlinks=true,
    linkcolor=blue,
    citecolor=green,
    filecolor=magenta,
    urlcolor=cyan
}

\begin{document}
%
\title{ContribChain: A Stress-Balanced Blockchain Sharding Protocol with Node Contribution Awareness}

\author{
\IEEEauthorblockN{
Xinpeng Huang\IEEEauthorrefmark{1}\IEEEauthorrefmark{2}\IEEEauthorrefmark{3},
Wanqing Jie\IEEEauthorrefmark{1}\IEEEauthorrefmark{2}\IEEEauthorrefmark{3},
Shiwen Zhang\IEEEauthorrefmark{1}\IEEEauthorrefmark{2},
Haofu Yang\IEEEauthorrefmark{1}\IEEEauthorrefmark{2}, 
Wangjie Qiu\IEEEauthorrefmark{1}\IEEEauthorrefmark{2}\IEEEauthorrefmark{3},
Qinnan Zhang\IEEEauthorrefmark{1}\IEEEauthorrefmark{2},\\
Huawei Huang\IEEEauthorrefmark{4},
Zehui Xiong\IEEEauthorrefmark{5},
Shaoting Tang\IEEEauthorrefmark{1}\IEEEauthorrefmark{2}\IEEEauthorrefmark{3},
Hongwei Zheng\IEEEauthorrefmark{6},
and Zhiming Zheng\IEEEauthorrefmark{1}\IEEEauthorrefmark{2}\IEEEauthorrefmark{3}
\IEEEauthorblockA{\IEEEauthorrefmark{1}Institute of Artificial Intelligence, Beihang University, Beijing, China}
\IEEEauthorblockA{\IEEEauthorrefmark{2}Beijing Advanced Innovation Center for Future Blockchain
  and Privacy Computing, Beihang University, Beijing, China}
\IEEEauthorblockA{\IEEEauthorrefmark{3}Zhongguancun Laboratory, Beijing, China}
\IEEEauthorblockA{\IEEEauthorrefmark{4}School of Software Engineering, Sun Yat-Sen University, China}
\IEEEauthorblockA{\IEEEauthorrefmark{5}Singapore University of Technology and Design, Singapore}
\IEEEauthorblockA{\IEEEauthorrefmark{6}Beijing Academy of Blockchain and Edge Computing (BABEC), Beijing, China}
\\\{huangxp, wanqingjie, zhangshiwen\}@buaa.edu.cn, \{2113824\}@mail.nankai.edu.cn, \{wangjieqiu, zhangqn\}@buaa.edu.cn, 
\\\{huanghw28\}@sysu.edu.cn, \{zehui\_xiong\}@sutd.edu.sg, \{tangshaoting\}@buaa.edu.cn, \{hwzheng, zzheng\}@pku.edu.cn}.
\thanks{* Wangjie Qiu and Qinnan Zhang are the \textbf{corresponding authors}.
}
}



\maketitle

\begin{abstract}
Existing blockchain sharding protocols have focused on eliminating imbalanced workload distributions. However, even with workload balance, disparities in processing capabilities can lead to differential stress among shards, resulting in transaction backlogs in certain shards. Therefore, achieving stress balance among shards in the dynamic and heterogeneous environment presents a significant challenge of blockchain sharding. In this paper, we propose ContribChain, a blockchain sharding protocol that can automatically be aware of node contributions to achieve stress balance. We calculate node contribution values based on the historical behavior to evaluate the performance and security of nodes. Furthermore, we propose node allocation algorithm NACV and account allocation algorithm P-Louvain, which both match shard performance with workload to achieve stress balance. Finally, we conduct extensive experiments to compare our work with state-of-the-art baselines based on real Ethereum transactions. The evaluation results show that P-Louvain reduces allocation execution time by 86\% and the cross-shard transaction ratio by 7.5\%. Meanwhile, ContribChain improves throughput by 35.8\% and reduces the cross-shard transaction ratio by 16\%.
\end{abstract}


%
\IEEEpeerreviewmaketitle

\section{Introduction}
With the rapid proliferation of distributed nodes constantly expanding, blockchain sharding has emerged as a promising a prominent technology used to improve the performance of blockchains \cite{Estuary,Elastico,Omniledger,Rapidchain,Monoxide,Brokerchain}. The core idea is to partition the whole blockchain network into many subnetworks, known as shards \cite{Elastico}. These shards process transactions in parallel, hence enhancing the transaction throughput of the blockchain. In addition, sharded blockchains undergo periodic reconfiguration of shards \cite{Omniledger,SGX-Sharding,Rapidchain} by altering the nodes within each shard to guarantee the security of whole blockchain network.

\begin{figure}[t] 
    \centering
    \includegraphics[width=0.9\linewidth]{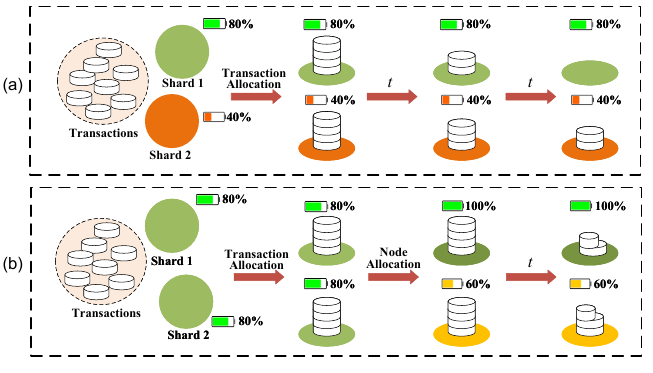}
    \caption{The stress imbalance issue in dynamic environments. Transaction backlogs occur in shards with low processing capabilities due to the lack of consideration of shard performance during transaction and node allocation.}
    \label{1-Problems}
\end{figure}

Currently, a significant amount of research has focused on reducing the ratio of cross-shard transactions and achieving load balance among shards to improve the performance of sharded blockchains \cite{Brokerchain,Achieving,Txallo,Estuary}. By leveraging the parallel advantages of sharding technology, these optimizations aim to increase system throughput. However, existing methods overlook the node composition within shards, focusing primarily on the transaction layer. When performance disparities exist among shards, even if their loads are balanced, their stress levels may not be. We define shard stress as the alignment between a shard’s transaction processing capacity and its workload. Fig. \ref{1-Problems} (a) and (b) highlight the impact of neglecting shard performance in transaction and node allocation, leading to transaction backlogs in shards with lower processing capabilities. This issue is referred to as stress imbalance among shards, occurring when a shard’s processing capacity and load do not align.

\textbf{Motivation.} Some studies \cite{Brokerchain,Achieving,Txallo} employ graph algorithms or community detection techniques for transaction allocation to reduce cross-shard transaction ratios and achieve load balancing among shards. However, in the account/balance model, the presence of popular accounts complicates the achievement of perfect load balance through transaction allocation alone. Recent work \cite{Estuary} employs state splitting and aggregation to allow users to hold accounts across multiple shards, while \cite{LB-Chain,AccountMigration} propose account migration mechanisms for rapid account reallocation. Nonetheless, these approaches are still focused on the transaction layer, leading to the two problems mentioned in Fig. \ref{1-Problems}.

Regarding shard reconfiguration, several studies \cite{SmartChain,SmartChain19,SmartChain20,SmartChain21} have applied reinforcement learning for adaptive node allocation. However, these methods have limitations: they require large amounts of data and long training time, and they do not address cross-shard transaction problems. Furthermore, these works lack consideration for shard load circumstances.

To summarize, current performance optimizations focus on either node or transaction allocation, without accounting for their combined impact. Additionally, these optimizations overlook the relationship between processing capabilities and load, resulting in stress imbalances among shards. To address this, we propose a stress-balanced sharding protocol that integrates both processing capabilities and load.

\textbf{Challenges.} In sharded blockchains, transaction processing within shards relies on consensus protocols among nodes. Consequently, the frequent ingress and egress of nodes lead to dynamic variations in a shard's transaction processing capability. Thus, accurately assessing node performance is crucial for estimating shard performance. However, developing a robust methodology to evaluate node performance is a significant challenge.

To tackle these challenges, we introduce ContribChain, a novel stress-balanced blockchain sharding protocol featuring dynamic node and account allocation algorithms based on node contribution values. The main \textbf{contributions} of this work are summarized as follows:

\begin{itemize}
    \item \textit{\textbf{A stress-balanced blockchain protocol (ContribChain):}} We propose dynamically updated node contribution values to comprehensively evaluate node performance and security. Additionally, our account and node allocation algorithms assess shard stress at both the node and transaction levels, ensuring stress balance across shards. 
    \item \textit{\textbf{Node allocation algorithm based on node contribution values (NACV) and performance-based account allocation algorithm (P-Louvain):}} NACV aims to balance the performance of allocated nodes and shard load, factoring in security (Section \ref{section III-C}). P-Louvain reduces the cross-shard transaction ratio while ensuring alignment between shard load and performance (Section \ref{section III-D}). 
    \item \textit{\textbf{System implementation:}} We implement P-Louvain and ContribChain on the open-source blockchain testbed \textit{BlockEmulator} \cite{Blockemulator}. Evaluation results demonstrate that, compared to state-of-the-art baselines, P-Louvain reduces allocation execution time by 86\% and the cross-shard transaction ratio by 7.5\%. Meanwhile, ContribChain improves throughput by 35.8\% and reduces the cross-shard transaction ratio by 16\%. ContribChain also shows superior stress balance and security performance.
\end{itemize}

The remainder of this paper is organized as follows: Section II reviews related work, Section III describes the protocol design, Section IV analyzes the security and other properties of ContribChain, Section V presents performance evaluation results, and Section VI concludes the paper.

\section{Related Work}
\subsection{Blockchain Sharding}
To address scalability limitations in blockchain systems, numerous sharding solutions have been proposed \cite{survey}. A review of key approaches is provided below. In 2016, Elastico \cite{Elastico} introduced the first blockchain sharding scheme, leveraging Proof of Work (PoW) for node selection and periodic reallocation. OmniLedger \cite{Omniledger} mitigates node storage burdens by employing full sharding, where each shard is responsible for storing non-overlapping ledger data. RapidChain \cite{Rapidchain} adopts full sharding and reconfigures nodes based on the Cuckoo rule \cite{Towards,Commensalcuckoo}. Later works \cite{Brokerchain,Achieving,Estuary} have expanded upon this reconfiguration model, focusing on reducing the costs associated with cross-shard transactions and ensuring more balanced workload distribution across shards in the context of full sharding.

\subsection{Node Allocation}
Node allocation is critical in blockchain sharding, involving both node assignment and reassignment. Regularly rotating nodes is essential to limit adversarial control, such as the 1/3 threshold in PBFT \cite{Practical}. Additionally, the allocation procedure must be unpredictable, unbiased and publicly verifiable \cite{Building}.

Node allocation strategies can be categorized into random, rule-based, and adaptive reconfiguration \cite{Building,SmartChain}. Pseudo-random number generators are used in \cite{Omniledger,SGX-Sharding} to select nodes for reassignment. Other methods \cite{Solida,Byzcoin} use criteria like activity level or chronological order. However, transferring nodes between shards incurs communication overhead due to ledger synchronization \cite{AccountMigration}. RapidChain \cite{Rapidchain} mitigates this by transferring only a subset of nodes at a time, known as the Cuckoo exchange mechanism. Despite the advances, the Cuckoo exchange mechanism lacks clear criteria for active nodes and does not consider shard performance, complicating post-reconfiguration performance assessments.

\subsection{Account Allocation}
Account allocation assigns accounts to specific shards, with each shard handling transactions related to its accounts. Early methods, such as \cite{Chainspace,Monoxide}, used hash-based approaches for transaction allocation, which could lead to uneven workloads and excessive cross-shard transactions.

To address these issues, \cite{Optchain,Challenges} model transactions or accounts as nodes in a graph, leveraging historical data and graph theory for allocation. Techniques in \cite{Txallo,Achieving} enhance community detection for more effective account assignment. BrokerChain \cite{Brokerchain} introduces brokers—users with accounts across multiple shards—to facilitate cross-shard transactions. Estuary \cite{Estuary} distributes accounts across shards, reducing cross-shard transactions. \cite{SPRING} employs Deep Reinforcement Learning (DRL) to optimize state placement. While these approaches enhance account allocation, they fail to consider performance disparities among shards, and thus do not achieve stress balance.

\section{ContribChain Protocol Design}
We introduce ContribChain, a protocol designed to assess node contributions and achieve stress balance through dynamic allocation of nodes and accounts. This section outlines the system model, workflow, and key algorithms.

\subsection{System Model And Workflow}
ContribChain operates on the account/balance model, consisting of two types of shards: one A-Shard and $K$ W-Shards. The protocol is designed to withstand a gradually adapting Byzantine adversary. The specific definitions are as follows:

\begin{figure*}[ht]
    \centering
    \includegraphics[width=0.95\textwidth]{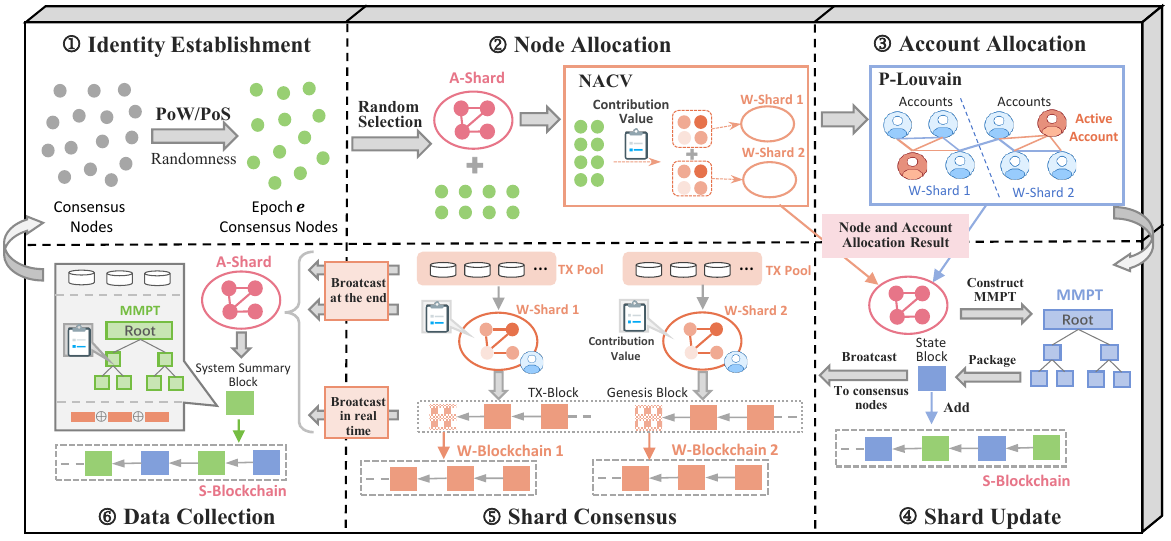}
    \caption{Workflow of ContribChain during epoch $e$. In ContribChain, Phase 3 is executed every $f$ epochs.}
    \label{System-Workflow}
\end{figure*}

\begin{itemize}

\item \textbf{W-Shard:} Each W-Shard processes client transactions in parallel. W-Shards use PBFT \cite{Practical} to achieve intra-shard consensus and ensures atomicity of cross-shard TXs through the Relay Transaction mechanism \cite{Monoxide}.

\item \textbf{A-Shard:} The A-Shard is responsible for allocating nodes and accounts to W-Shards and updating the global contribution values of all nodes.

\item \textbf{Node Contribution Values:} Node contribution values consist of security and performance contribution values, calculated based on historical behaviors. Stage contribution values are generated at each epoch to update the global contribution values.

\item \textbf{S-Blockchain:} The S-Blockchain in A-Shard records system-wide state changes, including node and account allocation results, and global node contribution values.

\item \textbf{W-Blockchain:} Each W-Shard maintains a W-Blockchain to record transaction processing results.

\end{itemize}

As illustrated in Fig. \ref{System-Workflow}, ContribChain operates in \textit{Epochs} \cite{Omniledger}, with the following workflow:

\textbf{\textit{Phase 1: Identity Establishment.}} To prevent Sybil attacks \cite{sybil}, nodes must obtain valid identities for the current epoch. Following previous methods \cite{Omniledger,Rapidchain,Brokerchain}, nodes use Verifiable Random Functions (VRF) to generate randomness, which is then used in a Proof of Work (PoW) protocol to solve a hash puzzle. PoW may also be replaced with Proof of Stake (PoS) for energy efficiency.

\textbf{\textit{Phase 2: Node Allocation.}} In Epoch 0, node identities are mapped to integers $\phi = [0, 2^{256} - 1]$ and divided into $K+1$ intervals. Nodes in the $(K+1)$th interval join the A-Shard, while others join the W-Shards \cite{Estuary}. From Epoch 1 onwards, the same method is first used to determine nodes in A-Shard. A-Shard then runs node allocation algorithm based on node contribution
value (NACV) (Section \ref{section III-C}) to update the nodes in W-Shards.

\textbf{\textit{Phase 3: Account Allocation.}} Every $f$ epochs, the A-Shard runs the P-Louvain algorithm (Section \ref{section III-D}) to allocate accounts to W-Shards.

\textbf{\textit{Phase 4: Shard Update.}} The A-Shard packages node and account allocation results into a State Block. After consensus, this block is added to the S-Blockchain and broadcast to all consensus nodes. W-Shards update nodes and accounts based on this block and generate the Genesis Block for the current epoch (Section \ref{section III-E}).

\textbf{\textit{Phase 5: Shard Consensus.}} W-Shards aggregate transactions into TX-Blocks. The leader of each W-Shard adds committed TX-Blocks to its W-Blockchain and sends them to the A-Shard in real-time. Then the leader records nodes' voting behaviors and block submission results.

\textbf{\textit{Phase 6: Data Collection.}} Each W-Shard calculates the stage contribution values of nodes (Section \ref{section III-B}) and packages them with pending transactions ($TX_{pending}$) into a Shard Summary Block. After consensus within the W-Shard, this block is sent to the A-Shard, which updates global contribution values. The contribution values, $TX_{pending}$, and other data are aggregated into a System Summary Block and added to the S-Blockchain (Section \ref{section III-E}).

\subsection{Node contribution values}\label{section III-B}
Node contribution values consist of performance contributions and security contributions, calculated based on voting behavior, node identity, and block submission results. At each epoch, nodes will have stage contribution values used to update the global node contribution values.

The stage performance contribution value $\Delta p_e(n_i)$ quantifies the TPS (Transactions Per Second) contributed by node $i$ during Epoch $e$. For successful block submissions,  the correct behavior is defined as a \textit{yes} vote, and the wrong behavior as a \textit{no} vote. For failed block submissions, the correct behavior is defined as a \textit{no} vote, and the wrong behavior as a \textit{yes} vote. The formula for $\Delta p_e(n_i)$ is:
\begin{equation}\label{equation-1}
\begin{split}
\Delta p_e(n_i) = \frac{\sum_{j=1}^{N_{\text{ALL}}} \left[ \frac{TX_j}{n_{R_j}} e_{i,j} 
- \delta_j \frac{TX_j}{n - n_{R_j}} (1 - e_{i,j}) \right]}{t_e}, 
\end{split}
\end{equation}

\begin{equation}\label{equation-2}
\delta_j = \begin{cases} 
0, & \text{if block $j$ submission succeeded;} \\ 
1, & \text{if block $j$ submission failed.}
\end{cases}
\end{equation}

\begin{equation}\label{equation-3}
e_{ij} = \begin{cases} 
1, & \text{if node } i \text{ behaved correctly;} \\ 
0, & \text{if node } i \text{ behaved incorrectly.}
\end{cases}
\end{equation}
Among them, $TX(i)$ represents the transactions contributed by node $i$. $t_e$ denotes the duration of Epoch $e$. $N_{ALL}$ denotes the total number of successful and failed block submissions in Epoch $e$. $n_{Rj}$ denotes the number of correctly behaving nodes during the consensus for block $j$. $TX_j$ denotes the transaction number in block $j$.

The stage security contribution value, $\Delta s_e(n_i)$, reflects the security performance of node $i$ in Epoch $e$, and is computed as:
\begin{equation}\label{equation-4}
\Delta s_e(n_i) = \frac{\mu \cdot \left( \lambda \cdot N_{MRi} + N_{FRi} \right) - \theta \cdot \left( \lambda \cdot N_{MWi} + N_{FWi} \right)}{\lambda \cdot \left( N_{MRi} + N_{FRi} \right) + N_{MWi} + N_{FWi}}.
\end{equation}
Among them, $N_{MRi}$ and $N_{MWi}$ are the numbers of correct and incorrect behaviors by node $i$ as a leader, and $N_{FRi}$ and $N_{FWi}$ are the numbers of correct and incorrect behaviors by node $i$ as a follower. $\mu$ and $\theta$ are the reward and penalty weights, respectively. $\lambda$ is the leader’s weight factor, set greater than 1 due to the higher impact of leader errors. From Eq. (\ref{equation-4}), it can be seen that $\Delta s_e(n_i) \in [-\theta, \mu]$. Further analysis of Eq. (\ref{equation-1}) and Eq. (\ref{equation-4}) is provided in Section \ref{section IV-A}. 

Global node contribution values are updated as follows:
\begin{equation}\label{equation-5}
s_e(n_i) = \alpha \cdot s_{e-1}(n_i) + (1 - \alpha) \cdot \Delta s_e(n_i), 
\end{equation}
\begin{equation}\label{equation-6}
p_e(n_i) = \alpha \cdot p_{e-1}(n_i) + (1 - \alpha) \cdot \Delta p_e(n_i), 
\end{equation}
where \( s_{e-1}(n_i) \) and \( p_{e-1}(n_i) \) are the security and performance global contribution values of node \( i \) at the end of Epoch \( e-1 \). \( \alpha \in [0, 1] \) is the retention factor, generally set to be greater than 0.5. Note that nodes in the A-Shard only update security contribution values, as they do not participate in transaction processing.

\subsection{The NACV Algorithm}\label{section III-C}

The A-Shard employs the Node Allocation Algorithm based on Node Contribution Values (NACV) to allocate nodes to W-Shards. The algorithm proceeds as follows:

\textbf{Step 1: Information Collection.} Obtain the security contribution value $s(n)$, performance contribution value $p(n)$ of each node, and the current node-shard mapping $M$ from the State Block. Calculate the security value $s(S)$, performance value $p(S)$, and estimated processing time $t(S)$ of each shard based on the workloads. Sort W-Shards twice by $s(S)$ and $t(S)$.

\textbf{Step 2: Handling New Nodes.} For new nodes without contribution values, assign the average security and performance contributions $\overline{p_{all}}$ and $\overline{s_{all}}$, respectively. For each node $n$ not present in the previous epoch, perform the following: If $s(n) \geq \overline{s_{all}}$, add $n$ to $\mathop{\arg\min}\limits_{S} s(S)$. If $p(n) \geq \overline{p_{all}}$, add $n$ to $\mathop{\arg\max}\limits_{S} t(S)$. Otherwise, add $n$ to $\mathop{\arg\max}\limits_{S} s(S)$. Update $s(S)$, $p(S)$, $t(S)$, $M$ and resort the shards.

\textbf{Step 3: Security Adjustment.} Set the initial and threshold values for the number of iterations, shard security variance and the output result of node-shard mapping:  \( I \), \( I_{thre} \), \( Var_{r}^{s} = \Var(s(S)) \), \( Var_{thre}^{s} \), \( M_{r} = M \). Iterate the following steps until \( Var_{r}^{s} \leq Var_{thre}^{s} \) or \( I \geq I_{thre} \):
(a) Sort shards by $s(S)$.
(b) Randomly select $n_i$ satisfying $s(n_i) \leq \overline{s_{all}}$ and $n_i \in \mathop{\arg\min}\limits_{S} s(S)$. Randomly select $n_j$ satisfying $s(n_j) \geq \overline{s_{all}}$ and $n_j \in \mathop{\arg\max}\limits_{S} s(S)$.
(c) Swap $n_i$ and $n_j$, update $Var(s(S)), M, I$ and resort shards by $s(S)$.
(d) If $Var_{r}^{s} < Var(s(S))$, update $Var_{r}^{s}$ and $M_{r}$.

\textbf{Step 4: Performance Adjustment.} Set the initial and threshold values for shard processing time variance and reset the number of iterations: $Var_{r}^{t}=\Var(t(S))$,  $Var_{thre}^{t}$, $I=0$. Iterate the following steps until $Var_{r}^{t} \leq Var_{thre}^{t}$ or $I \geq I_{thre}$:
(a) Calculate $Var(t(S))$.
(b) Randomly select $n_i$ satisfying $p(n_i) \geq \overline{p_{all}}$ and $n_i \in \mathop{\arg\min}\limits_{S} t(S)$. Randomly select $n_j$ satisfying $p(n_j) \leq \overline{p_{all}}$ and $n_j \in \mathop{\arg\max}\limits_{S} t(S)$.
(c) If not found, select $n_i$ satisfying $\arg \max_{n_i \in \mathop{\arg \min}\limits_{S} t(S)} p(n_i)$. Randomly select $n_j$ satisfying $p(n_j)<p(n_i)$ and $n_j \in \mathop{\arg\max}\limits_{S} t(S)$. 
(d) If not found, select $n_i$ satisfying $\arg \max_{n_i \notin \mathop{\arg \max}\limits_{S} t(S)} p(n_i)$. Randomly select $n_j$ satisfying  $p(n_j)<p(n_i)$ and $n_j \in \mathop{\arg\max}\limits_{S} t(S)$.
(e) Swap $n_i$ and $n_j$ if suitable nodes are found. Otherwise, abandon node allocation and trigger account allocation.
(f) Update $Var(t(S)), Var(s(S)), M_{r}, I$. If $Var(t(S))<Var_{r}^{t}  $ and $Var(s(S))<Var_{thre}^{s}$, update $Var_{r}^{t}, M_{r}$.

The algorithm terminates with the output of the updated node-shard mapping $M_r$. The time complexity is $O(N' \cdot K + I_{\text{thre}} \cdot K \log K)$, where $N'$ is the number of new nodes, $I_{\text{thre}}$ is the iteration threshold, and $K$ is the number of shards.

\subsection{The P-Louvain Algorithm}\label{section III-D}
We propose the P-Louvain algorithm for account allocation, which extends the community detection algorithm \textit{Louvain} \cite{louvain} by considering the performance of shards. The leader in A-Shard executes P-Louvain to minimize disparities in processing time across W-Shards. As shown in Algorithm \ref{algorithm}, we define the transaction graph as $\mathcal{G} = (\mathcal{V}, \mathcal{E})$, the number of shards as $K$, performance values of shards as $\mathcal{P} = \{p_1, p_2, \ldots, p_K\}$ and the accounts lists of shards as $\mathcal{V} = \{V_1, \ldots, V_K\}$. $p_K$ is estimated by summing the performance contribution values of nodes in shard $K$, as demonstrated in Section \ref{section IV-A}. 

\textbf{Initialization Phase:} We employ \textit{Louvain} on $\mathcal{G}$ to iteratively divide accounts into communities. Typically, the number of communities exceeds the number of shards.

\textbf{Community Movement Phase:} We sort communities by size and sort shards by $\mathcal{P}$. The first $K$ communities are added to the corresponding shards (Lines 6-9). The remaining communities are processed successively in lines 10-14. Each community is assigned to the shard with the minimum processing time, and the processing time of shards $\mathcal{T}$ is then updated. Adding this phase can lead to faster convergence of the algorithm.

\textbf{Node Movement Phase:} We initialize the check flag of each account to true. Subsequently, lines 17-32 loop through the account $v_i$ whose check flag is true. First we get the list of shards where the account's neighboring accounts on $\mathcal{G}$ are located. Then for each $S$ in the list, we calculate the new processing time for the shards that $v_i$ came from and went to if moving $v_i$ to $S$. Then we calculate the reduction of the maximum value of the processing time for both shards, recording the maximum reduction $R_{\text{max}}$ and the corresponding shard $S_{\text{max}}$. Subsequently, we move $v_i$ to $S_{\text{max}}$ and update $\mathcal{T}$, $\mathcal{V}$. The check flag of $v_i$ and its neighbors are set to false and true respectively (Lines 28-31). Finally, when there are no accounts to check, P-Louvain outputs $\mathcal{V}$. 

The time complexity of each phase is $O(m \log n)$, $O(l \log l + lK)$ and $O(mK)$. Among them, $m$ is the number of edges, $n$ is the number of accounts and $l$ is the number of communities. The total time complexity is $O(m \log n + l \log l + lK + mK)$.

We develop a verification procedure for other nodes to verify the result $\mathcal{V}$. If any boundary account can be moved to reduce shard processing times, the verification fails. If no moves are possible, the verification succeeds.


\begin{algorithm}[t]
\DontPrintSemicolon
\fontsize{10pt}{10.8pt}\selectfont 
  \SetAlgoLined
  
  \KwIn{
    $\mathcal{G} = (\mathcal{V}, \mathcal{E})$, $K$, $\mathcal{P} = \{p_1, p_2, \ldots, p_K\}$.
  }
  \KwOut{
    $\mathcal{V} = \{V_1, \ldots, V_K\}$.
  }

  \textcolor{darkgreen}{// 0. Initialization Phase.} \;
  $\{c_1, c_2, \ldots, c_K, \ldots, c_l\} = \text{Louvain}(\mathcal{G})$.\;

  \textcolor{darkgreen}{// 1. Community Movement Phase.} \;
  Sort communities by size $\rightarrow \{c_1, c_2, \ldots, c_K, \ldots, c_l\}$.\;
  Sort shards by $\mathcal{P} \rightarrow \{S_1, S_2, \ldots, S_K\}$.\;
  \For{$i = 1$ \text{ to } $K$}{
    Move community $c_i$ to shard $S_i$.\;
    Update shard processing time $\mathcal{T}=\{t_1, t_2, \ldots, t_K\}$.\;
  }
  \For{$i = K + 1$  \text{ to }  $l$}{
    Find shard $S_{\text{minTime}}$with the minimum $t$.\;
    Move community $c_i$ to shard $S_{\text{minTime}}$.\;
    Update $\mathcal{T}$.\;
  }

  \textcolor{darkgreen}{// 2. Account Movement Phase.} \;
  $\text{needCheck}[v_i] = \text{true}$, each $v_i \in \mathcal{V}$.\;
  \While{$v_i \in \mathcal{V}$ and $\text{needCheck}[v_i] == \text{true}$}{
    Get neighbor’s non-repetitive shard list $\mathcal{NS}_{v_i} = \{\mathcal{NS}_{v_i}^1, \mathcal{NS}_{v_i}^2, \ldots, \mathcal{NS}_{v_i}^j\}$.\;
    $R_{\text{max}} \leftarrow 0$, $S_{\text{max}}\leftarrow$ the  shard  where  $v_i$  is located.\;
    \For{$S \in \mathcal{NS}_{v_i}$}{
      \If{$v_i \notin S$}{
        Calculate two shards’ new processing time $t_{\text{from}}', t_{\text{to}}'$, if move $v_i$ to $S$.\;
        \If{$\max(t_{\text{from}}', t_{\text{to}}') - \max(t_{\text{from}}, t_{\text{to}}) > R_{\text{max}}$}{
          Update $R_{\text{max}}, S_{\text{max}}$.\;
        }
      }
    }
    Move $v_i$ to $S_{\text{max}}$.\;
    Update $\mathcal{T}$, $\mathcal{V}$.\;
    $\text{needCheck}[u] = \text{true}$, each neighbor $u$ of $v_i$.\;
    $\text{needCheck}[v_i] = \text{false}$.\;
  }
  \Return $\mathcal{V} = \{V_1, V_2, \ldots, V_K\}$.\;

  \caption{The P-Louvain Algorithm.}
\label{algorithm}
\end{algorithm}

\subsection{System state update mechanism}\label{section III-E}
This subsection describes the state update mechanism during epoch switches. The leader of each W-Shard calculates the stage contribution values for Epoch $e$ using Eq. (\ref{equation-1}) and Eq. (\ref{equation-4}) and A-Shard packages them with $TX_{pending}$ into the Shard Summary Block. Two Modified Merkle Patricia Trees (MMPT) \cite{MPT} are created, with their roots added to the header of the Shard Summary Block. 

\begin{figure}[t]
    \centering
    \includegraphics[width=0.9\linewidth]{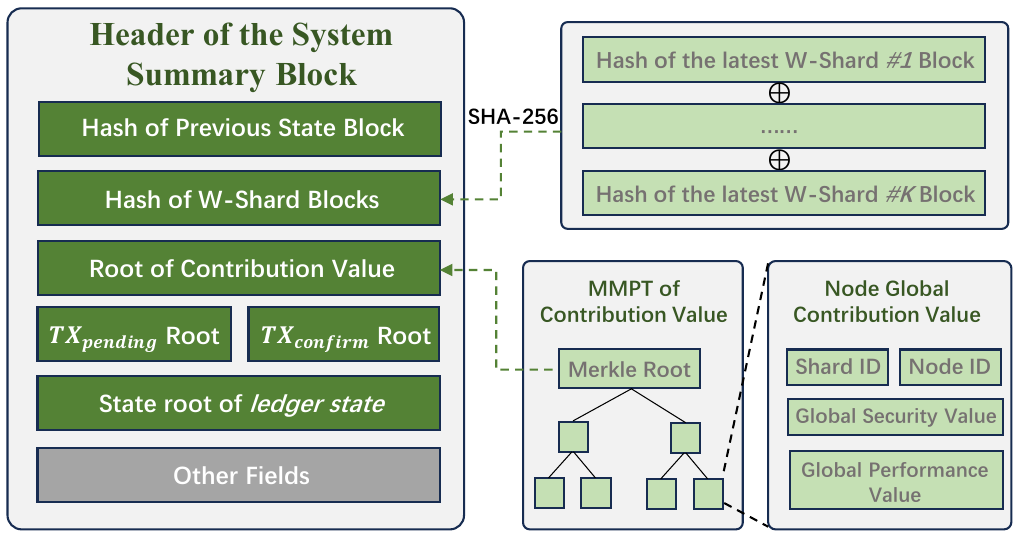}
    \caption{Data Structure of the System Summary Block.}
    \label{systemSummaryBlock}
\end{figure}

A-Shard generates the System Summary Block for Epoch $e$, storing the hash of the latest TX-blocks from W-Shards in the block header, as illustrated in Fig. \ref{systemSummaryBlock}. Additionally, A-Shard constructs the updated global contribution values into MMPT and adds the root of MMPT, the root of confirmed transactions \(TX_{confirm}\), the root of \(TX_{pending}\), and other fields into the header of the block. Other fields contain information such as block height. After consensus, the block is submitted to the S-Blockchain.

\begin{figure}[t]
    \centering
    \includegraphics[width=0.9\linewidth]{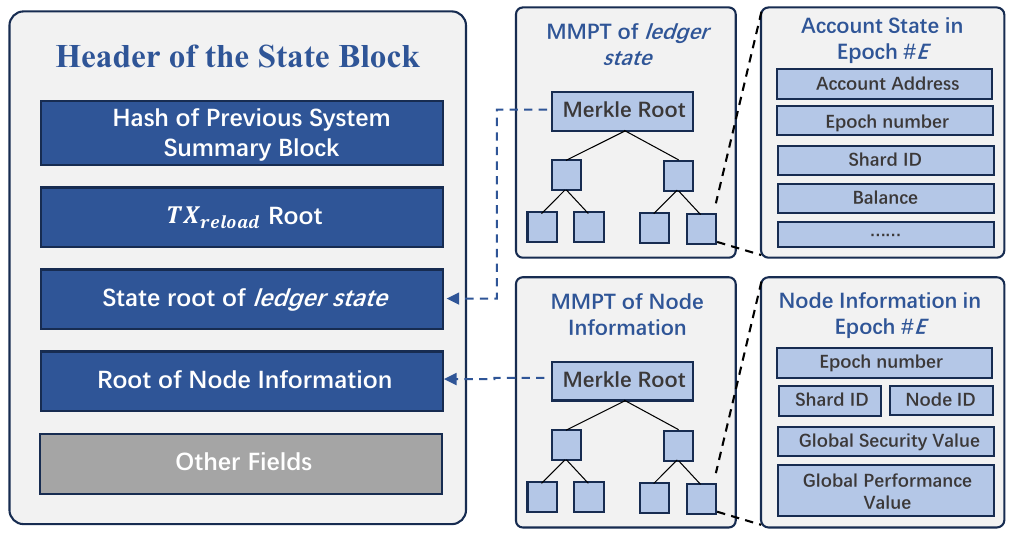}
    \caption{Data Structure of the State Block.}
    \label{stateBlock}
\end{figure}

Upon entering Epoch $e+1$, after node and account allocation, A-Shard constructs two MMPTs with the updated information of nodes and accounts. The roots of the two MMPTs are stored in the header of the State Block, as illustrated in Fig. \ref{stateBlock}. Depending on the result of P-Louvain, $TX_{pending}$ in Epoch $e$ are organized as $TX_{reload}$, with their root included in the header of this block. After consensus, the State Block is submitted to S-Blockchain and broadcast. Nodes determine their respective W-Shards and $TX_{reload}$ based on this block. Each W-Shard generates a Genesis Block to achieve consensus on the updated state, and begins processing transactions.


\section{Analysis}
This section presents the rationale behind the node contribution values, shard security, and ledger security.  
\subsection{The Reasonableness of the Node Contribution Values}\label{section IV-A}
We analyze the reasonableness of the security and performance contribution values to ensure they accurately reflect the nodes' contribution.

\textit{\textbf{Theorem 1.} The security contribution value of an honest node remains constant, while that of a malicious node decreases.}

\textit{\textbf{Proof}}: Consider node $i$ as a follower during Epoch $e$. Eq. (\ref{equation-4}) can be organized as:
\begin{align}\label{equation-7}
 \Delta s_e(n_i) &= \mu \cdot \frac{N_{FRi}}{N_{FRi} + N_{FWi}} - \theta \cdot \frac{N_{FWi}}{N_{FRi} + N_{FWi}} \notag \\
 &= \mu \cdot f_{Ri} - \theta \cdot f_{Wi},
\end{align}
where $f_{Ri}$ and $f_{Wi}$ represent the ratios of accurate and inaccurate behaviors of node $i$ during Epoch $e$, respectively. Consequently, $f_{Ri} + f_{Wi} = 1$. Given that $\theta > \mu$, an increase in $f_{Wi}$ results in a decrease in $\Delta s_e(n_i)$, meaning $s_e(n_i) < s_{e-1}(n_i)$ for malicious nodes.

In PBFT, the leader moves to the next step after collecting commit messages from 2/3 of nodes. Therefore, even if all nodes are honest, there will still be 1/3 of nodes considered to be behaving incorrectly. In this case, $s_e(n_i) \in \left[ \frac{2\mu - \theta}{3} - \epsilon, \frac{2\mu - \theta}{3} + \epsilon \right]$, where $\epsilon$ represents a small fluctuation. Thus, we set $\theta > \mu \geq \frac{\theta}{2}$ to ensure that the security contribution value of an honest node is greater than or equal to 0. And we can derive the condition for $\Delta s_e(n_i) < 0$ from Eq. (\ref{equation-7}):

\begin{equation}
\Delta s_e(n_i) < 0 \iff f_{Wi} > \frac{\mu}{\mu + \theta}. 
\end{equation}

Therefore, by changing $\mu$ and $\theta$, the system's tolerance for malicious behaviors can be adjusted. If $\mu = \frac{\theta}{2}$, then $s_e(n_i) \in \left[ -\epsilon, \epsilon \right]$, $\Delta s_e(n_i) < 0 \iff f_{Wi} > \frac{1}{3}$, which represents the strictest condition.

\textit{\textbf{Theorem 2.} The sum of the stage performance contribution values of the nodes within a shard during an epoch equals the shard's TPS.}

\textit{\textbf{Proof}:} Let $N_{\text{ALL}}$ denote the number of blocks in shard $S$ during Epoch $e$, with $TX_j$ being the number of transactions in block $j$ and $t_e$ the duration of Epoch $e$. Combining Eq. (\ref{equation-2}), the TPS of shard $S$ during Epoch $e$ can be expressed as:
\begin{align}\label{equation-9}
TPS_e^S & = \frac{1}{t_e} \sum_{j=1}^{N_{\text{ALL}}} (TX_j - \delta_j \cdot TX_j). 
\end{align}

Based on the reward rules (Section \ref{section III-B}), the transaction contribution $TX(n_i)$ for node $i$ is calculated. If block $j$ is successfully submitted, $TX(n_i)$ of each node that behaved correctly is \(\frac{TX_j}{n_{R_j}}\), and there are no penalties. If the submission fails, $TX(n_i)$ of each node that behaved correctly is \(\frac{TX_j}{n_{R_j}}\), while that of each node that behaved incorrectly is \(\frac{- TX_j}{n - n_{R_j}}\).

Combining Eq. (\ref{equation-2}) and Eq. (\ref{equation-3}), we can calculate the stage contribution value $\Delta p_e(n_i)$ as follows:
\begin{equation}
\Delta p_e(n_i) = \frac{1}{t_e} \sum_{j=1}^{N_{\text{ALL}}} \left[ \frac{TX_j}{n_{R_j}} e_{i,j} 
- \delta_j \frac{TX_j}{n - n_{R_j}} (1 - e_{i,j}) \right]. 
\end{equation}

The total number of nodes in $S$ is represented as $n$. Then, the sum of all node contributions can be calculated as:

\begin{equation}
\begin{split}
\sum_{i=0}^{n-1} \Delta p_e(n_i) 
&= \frac{1}{t_e} \sum_{j=1}^{N_{\text{ALL}}} \sum_{i=0}^{n-1} \left[ \frac{TX_j}{n_{Rj}} e_{ij} 
- \frac{\delta_j \cdot TX_j}{n - n_{Rj}} (1 - e_{ij}) \right] \\
&= \frac{1}{t_e} \sum_{j=1}^{N_{\text{ALL}}} \left[ \frac{TX_j}{n_{Rj}} n_{Rj} 
- \frac{\delta_j \cdot TX_j}{n - n_{Rj}} (n - n_{Rj}) \right] \\
&= \frac{1}{t_e} \sum_{j=1}^{N_{\text{ALL}}} (TX_j - \delta_j \cdot TX_j).
\end{split}
\end{equation}

Combining Eq. (\ref{equation-9}), \textit{Theorem 2} concludes.

\subsection{Shard Security}

The A-Shard node selection method follows the random selection approach from \cite{Estuary}, \cite{Rapidchain}. Given that PBFT requires the proportion of malicious nodes to be less than 1/3, the failure probability of A-Shard can be derived from the cumulative hypergeometric distribution \cite{Building}, as expressed below:  
\begin{equation}
Pr[X\geq\lceil n_A/3\rceil]=\sum_{x=\lceil n_A/3\rceil}^{n_A} \frac{\binom{\rho n}x\binom{n(1-\rho)}{n_A-x}}{\binom n{n_A}}.
\label{e-7}
\end{equation}
Among them, $n$ is the total number of nodes, $n_A$ is the number of nodes in A-Shard, $X$ is the number of malicious nodes in A-Shard and $\rho$ is the proportion of malicious nodes in the system. Next, we analyze the security of W-Shard.

\textit{\textbf{Theorem 3.} The NACV algorithm is unpredictable and verifiable.}

\textbf{\textit{Proof:}} The random selection operation in NACV can be described as choosing a node that meets a specific condition from the set  $Set^{special}_w$ within a W-shard. The selection process is as follows: First, the randomness from the current epoch serves as the seed for a pseudo-random number generator to generate $Output\in \{1, 2, \ldots, |Set^{special}_w|\}$. The nodes in $Set^{special}_w$ are then sorted in ascending order by their addresses, and the node corresponding to $Output$ is selected. Since the same seed results in the same $Output$, all S-Shard nodes will select the same node, ensuring that NACV is deterministic and verifiable. Furthermore, the use of randomness guarantees the unpredictability of the selection outcome. 

\textit{\textbf {Theorem 4.} If A-Shard is secure, W-Shards will be secure. }

\textit{\textbf{Proof:}} According to \textit{\textbf{Theorem 1}}, a decrease in a node's security value indicates an increase in its malicious behavior, which may result from degraded network conditions, reduced computational capacity, or adversarial attacks. If A-Shard is secure, the NACV algorithm will distribute nodes with lower security contributions across W-Shards to minimize the variance in shard security. This ensures that the security of W-Shards is maintained. 

\subsection{Node Contribution and Ledger Security} 

Malicious nodes may attempt to manipulate node contribution values and account balances. In ContribChain, both stage and global contribution values are represented as MMPTs, with their roots placed at the block headers (Section \ref{section III-E}). These blocks are propagated throughout the system for consensus, allowing nodes to verify the integrity of contribution values using the Merkle path \cite{MPT}. Similarly, the correctness of account balances in the state blocks can be validated in the same manner \cite{Achieving}.

\begin{figure*}[ht]
    \centering
    \begin{subfigure}[b]{0.235\textwidth}
        \centering
        \includegraphics[width=\textwidth]{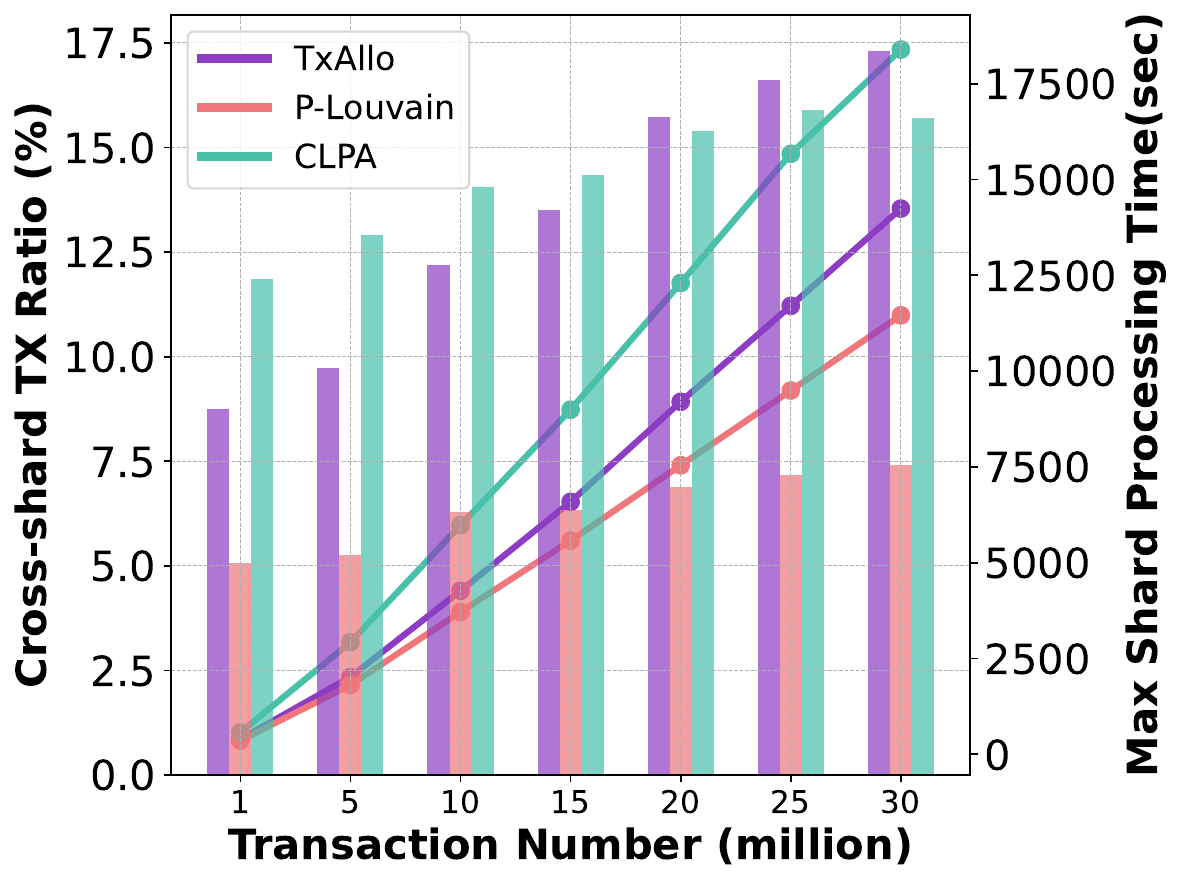} 
        \caption{Effect of allocation, $K$=4}
        \label{t-1}
    \end{subfigure}
    \hfill
    \begin{subfigure}[b]{0.24\textwidth}
        \centering
        \includegraphics[width=\textwidth]{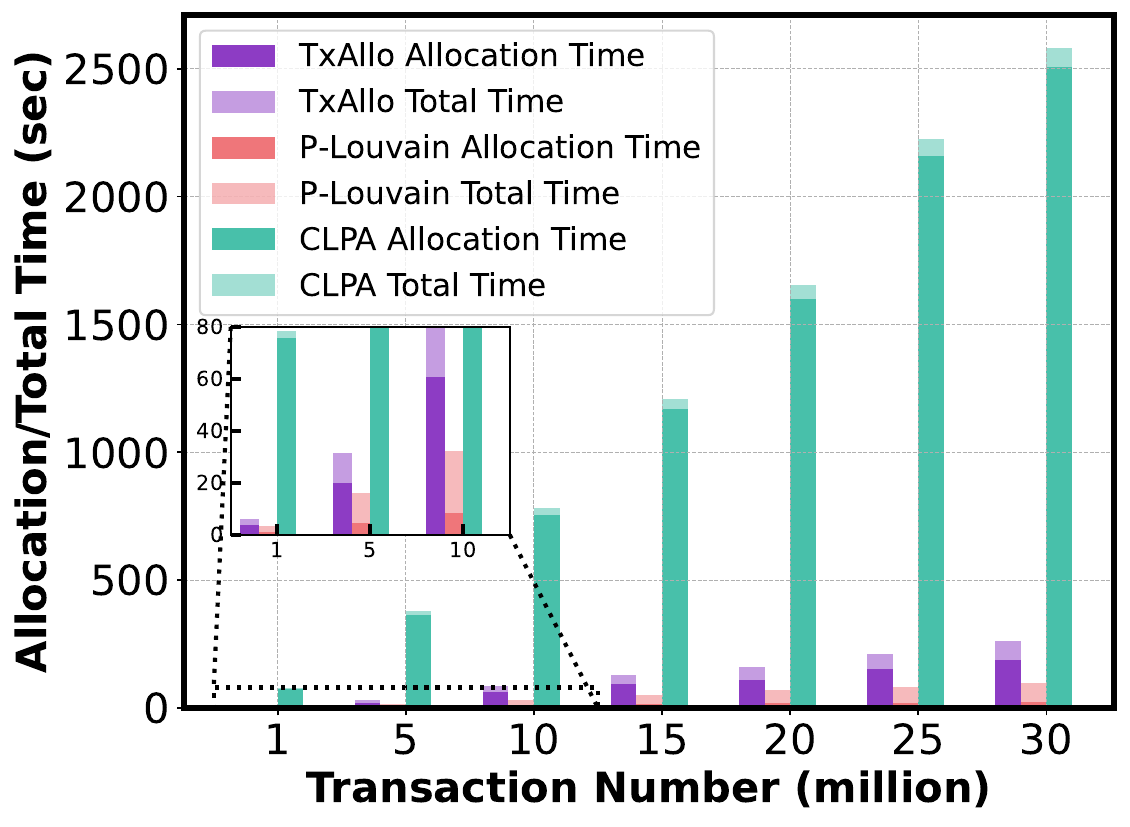} 
        \caption{Time consumption, $K$=4}
        \label{t-2}
    \end{subfigure}
    \hfill
    \begin{subfigure}[b]{0.24\textwidth}
        \centering
        \includegraphics[width=\textwidth]{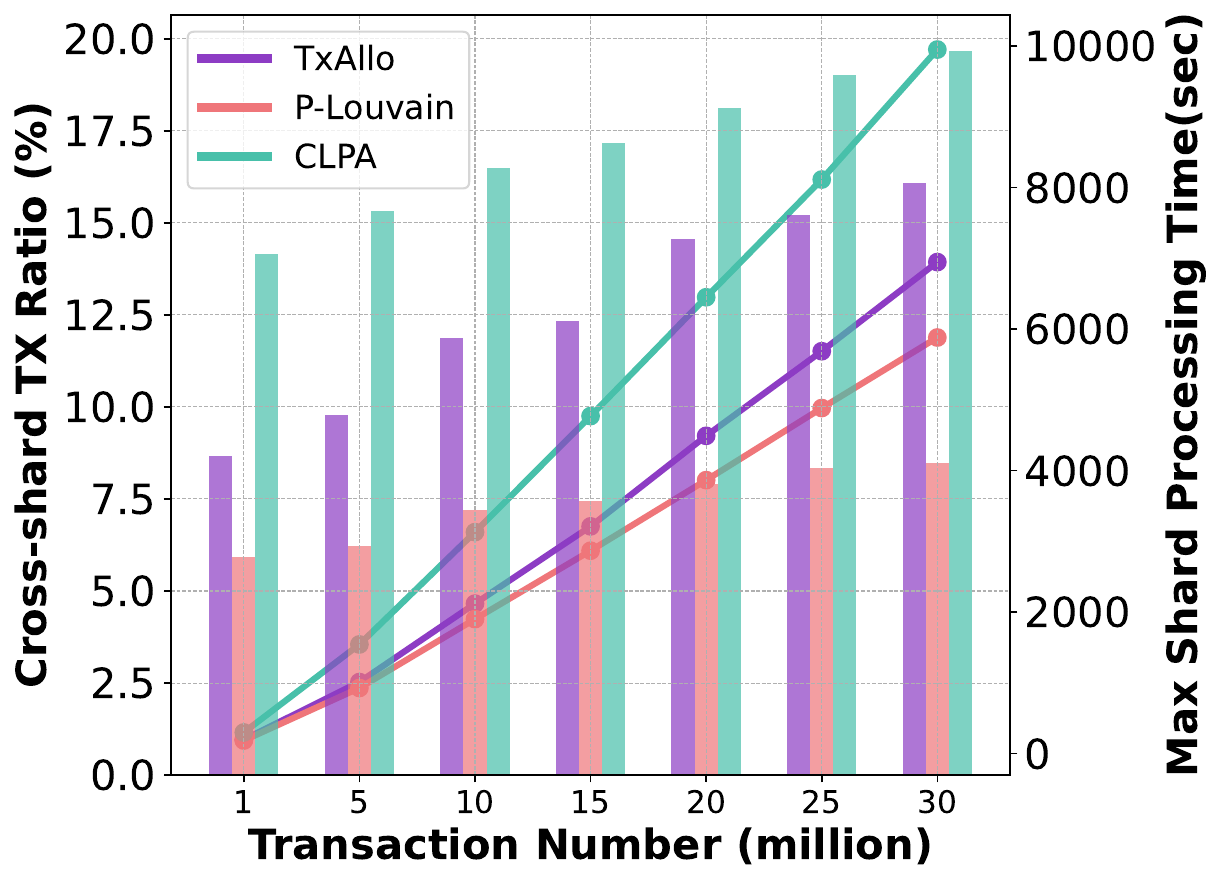} 
        \caption{Effect of allocation, $K$=8}
        \label{t-3}
    \end{subfigure}
    \hfill
    \begin{subfigure}[b]{0.24\textwidth}
        \centering
        \includegraphics[width=\textwidth]{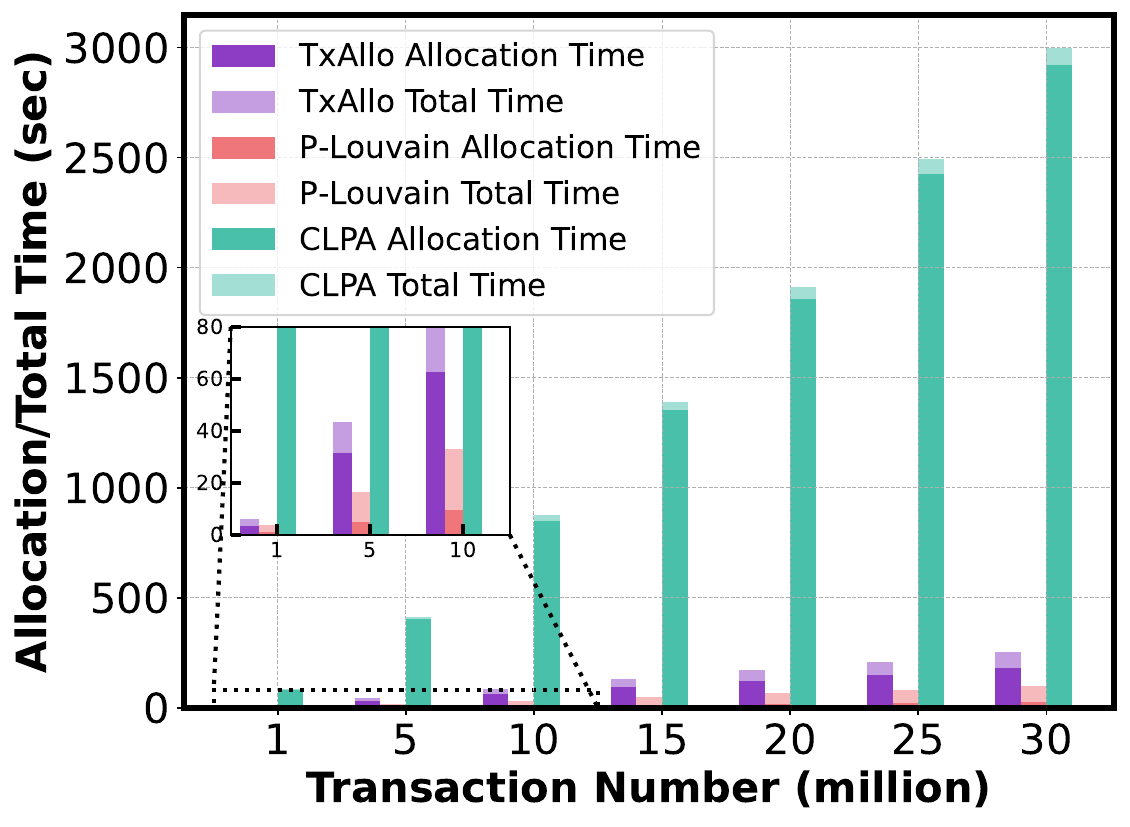} 
        \caption{Time consumption, $K$=8}
        \label{t-4}
    \end{subfigure}
    \caption{Performance of account allocation algorithms with varying $N_{TX}$. In (a) and (b), assume the TPS of shards are \{1000, 800, 800, 600\}. In (c) and (d), assume the TPS of shards are \{1000, 900, 900, 800, 800, 700, 700, 600\}.}
    \label{fig:load-tps}
\end{figure*}
\begin{figure*}[ht]
    \centering
    \begin{subfigure}[b]{0.24\textwidth}
        \centering
        \includegraphics[width=\textwidth]{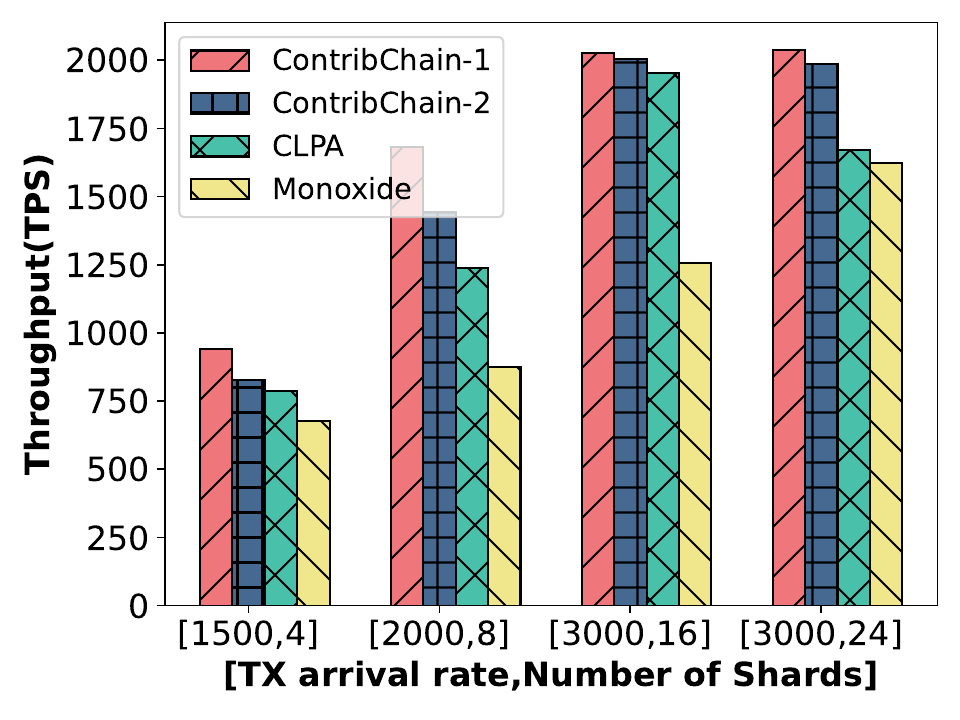} 
        \caption{TPS}
        \label{a-tps}
    \end{subfigure}
    \hfill
    \begin{subfigure}[b]{0.24\textwidth}
        \centering
        \includegraphics[width=\textwidth]{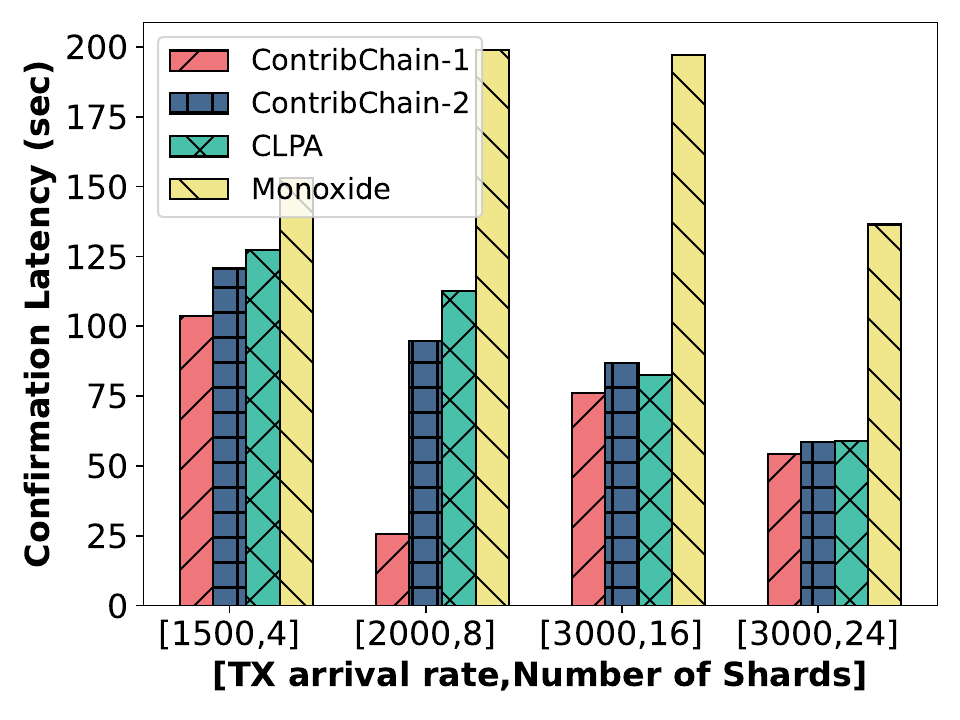} 
        \caption{Latency}
        \label{a-latency}
    \end{subfigure}
    \hfill
    \begin{subfigure}[b]{0.25\textwidth}
        \centering
        \includegraphics[width=\textwidth]{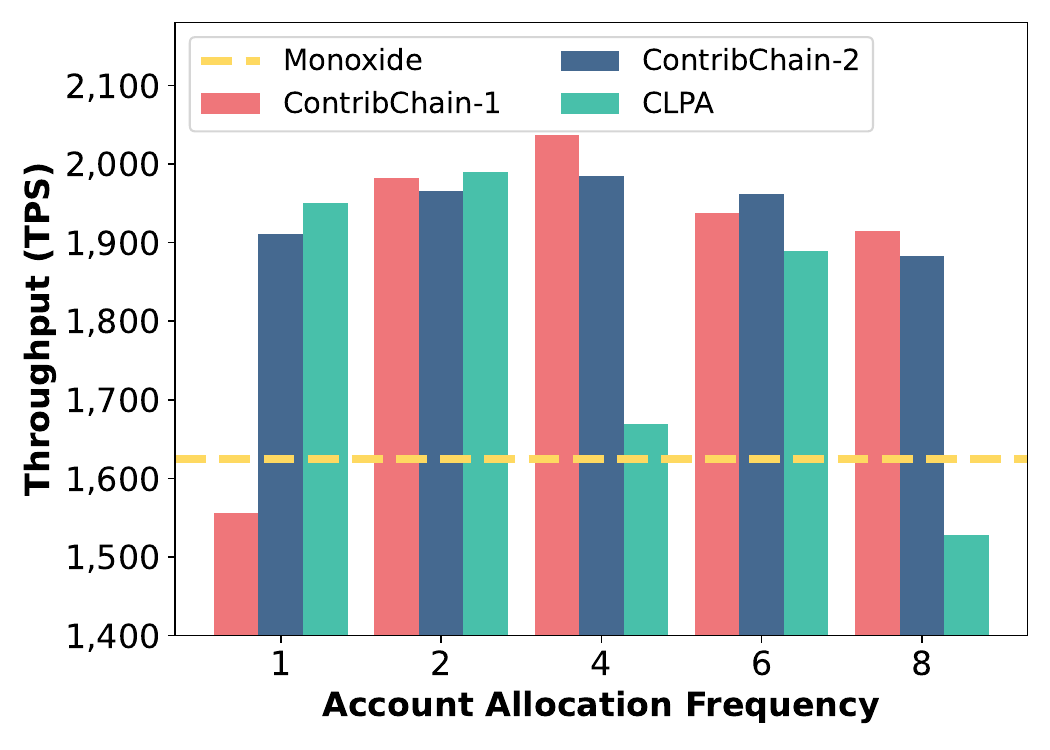} 
        \caption{TPS, $T_{NA}=80$s}
        \label{b-tps}
    \end{subfigure}
    \hfill
    \begin{subfigure}[b]{0.25\textwidth}
        \centering
        \includegraphics[width=\textwidth]{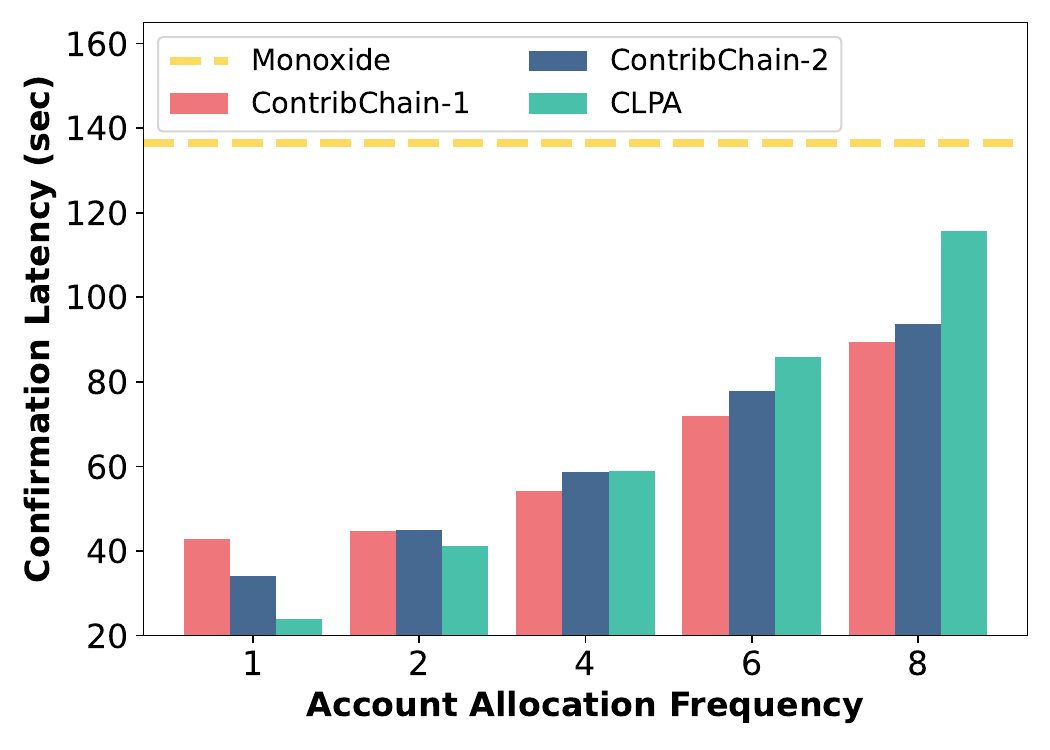} 
        \caption{Latency, $T_{NA}=80$s}
        \label{b-latency}
    \end{subfigure}
    \caption{TPS and transaction confirmation latency changes with TX arrival rate, $N_{TX}$ and account allocation frequency ($f$).}
    \label{a}
\end{figure*}

\begin{figure*}[ht]
    \centering
    \begin{subfigure}[b]{0.238\textwidth}
        \centering
        \includegraphics[width=\textwidth]{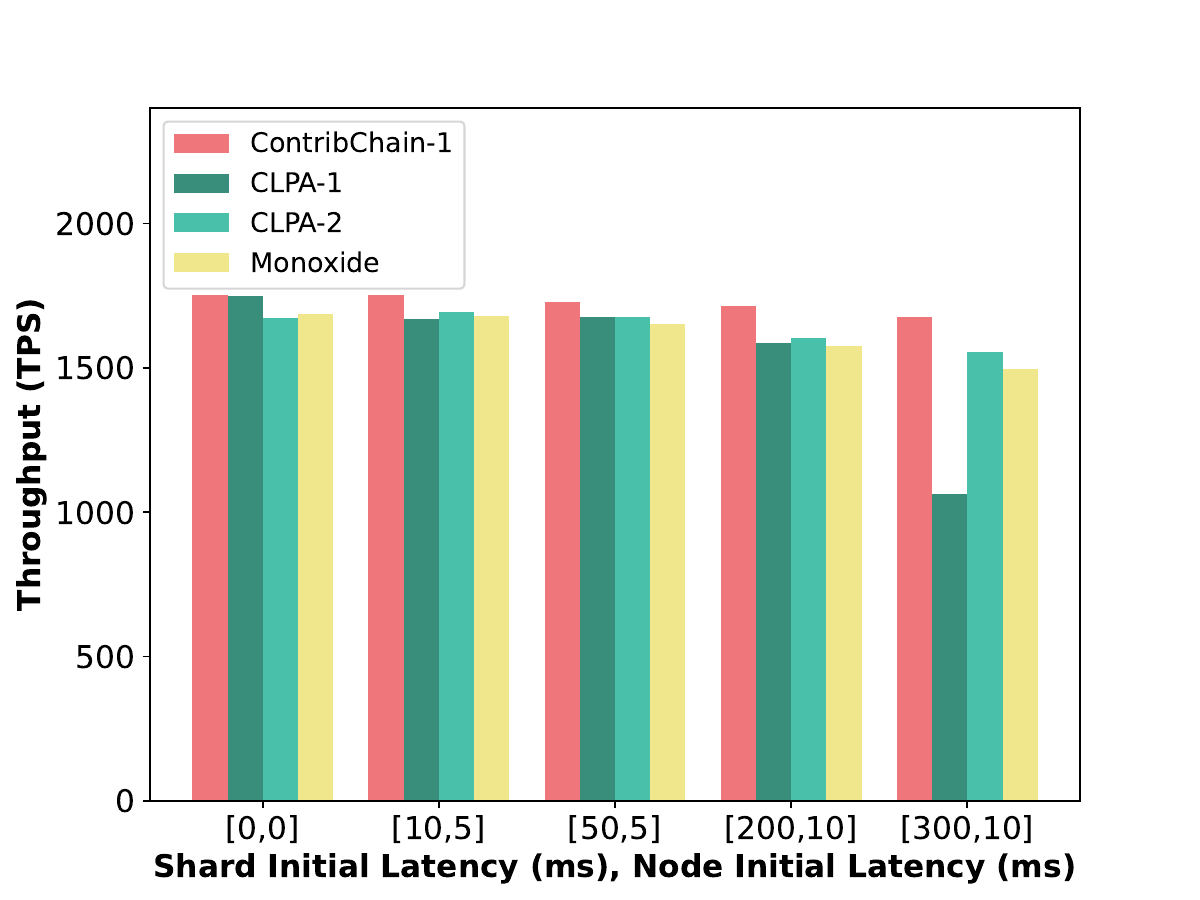} 
        \caption{TPS}
        \label{C-TPS}
    \end{subfigure}
    \hfill
    \begin{subfigure}[b]{0.215\textwidth}
        \centering
        \includegraphics[width=\textwidth]{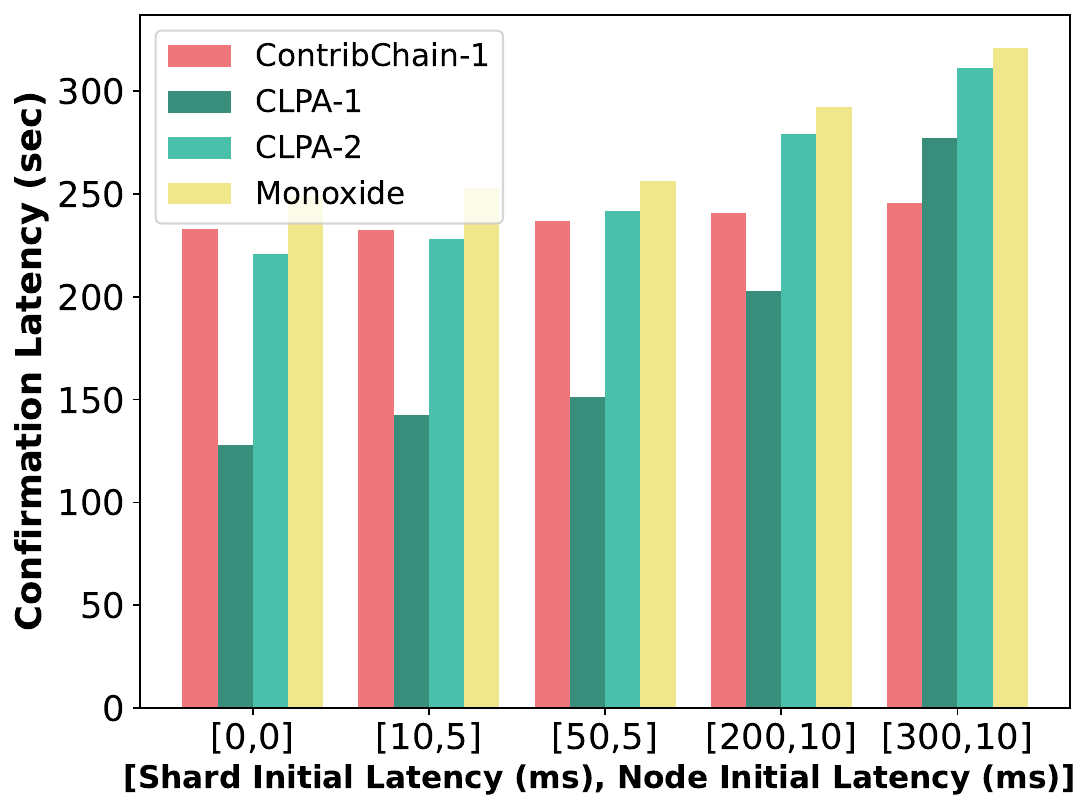} 
        \caption{Latency}
        \label{C-Latency}
    \end{subfigure}
    \hfill
    \begin{subfigure}[b]{0.25\textwidth}
        \centering
        \includegraphics[width=\textwidth,height=\dimexpr0.65\textwidth\relax]{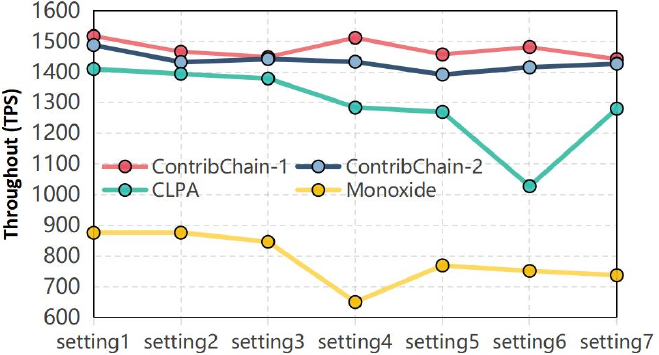}
        \caption{TPS}
        \label{D-TPS}
    \end{subfigure}
    \hfill
    \begin{subfigure}[b]{0.25\textwidth}
        \centering
        \includegraphics[width=\textwidth,height=\dimexpr0.65\textwidth\relax]{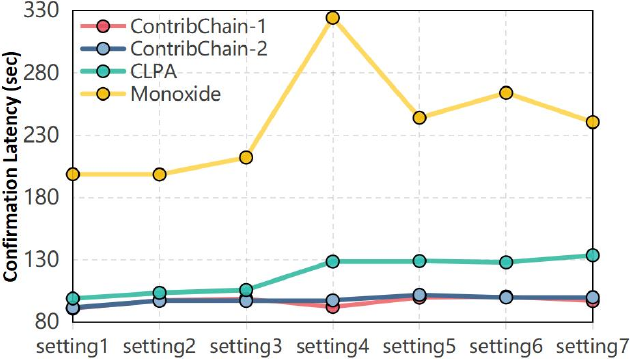} 
        \caption{Latency}
        \label{D-Latency}
    \end{subfigure}
    \caption{TPS and transaction confirmation latency changes with node delay settings or both node security and delay settings.}
    \label{b}
\end{figure*}
\section{EVALUATION}

\subsection{Experiment Settings}
We implemented P-Louvain and ContribChain on an open-sourced blockchain testbed \textit{BlockEmulator} \cite{Blockemulator}.

\textbf{Dataset:} We used historical Ethereum transactions \cite{ethereum} from block height 10,000,000 to 10,999,999 (from May 4, 2020, to October 6, 2020). The maximum number of transactions used reached 30 million.

\textbf{Transaction Processing:} Each block is set to contain a maximum of 2000 TXs. The block interval is set to 5 seconds. Transactions are sent to the system at a specific injection rate.

\textbf{Baselines:} P-Louvain was compared against TxAllo \cite{Txallo} and CLPA \cite{Achieving}. TxAllo is another account allocation algorithm based on \textit{Louvain}. In the system experiments, we set ContribChain-1 to use NACV and P-Louvain, and ContribChain-2 to use only P-Louvain. Then we compared them with CLPA and Monoxide \cite{Monoxide}, where Monoxide allocates accounts based on the first few digits of the account address.

\textbf{Other Settings:} The parameters $\mu$, $\theta$, $\lambda$, $\alpha$, and $\beta$ were set to 0.9, 1.5, 2, 0.7, and 2, respectively.

\begin{table}[h]
\centering
\caption{Node Safety and Delay Settings. $Num_{n_w}$ denotes the number of potential malicious nodes per shard, $P_{n_w}$ represents the probability of malicious behaviors, and ($L_s, L_n$) indicates the initial latency for shards and nodes (ms).}
\label{settings}
\begin{tabular}{|p{1cm}|p{2cm}|p{2cm}|p{2cm}|}
\hline
\textbf{Setting} & \textbf{$Num_{n_w}$} & \textbf{$P_{n_w}$} & \textbf{$L_s, L_n$} \\
\hline
1 & 1 & 5\% & 0, 0 \\
\hline
2 & 1 & 5\% & 10, 5 \\
\hline
3 & \{0, 1, 2\} & 5\% & 50, 5 \\
\hline
4 & \{0, 1, 2\} & 5\% & 200, 10 \\
\hline
5 & \{0, 1, 2\} & [20\%, 54\%] & 200, 10 \\
\hline
6 & \{0, 1, 2\} & 100\% & 300, 10 \\
\hline
7 & \{0, 2, 2\} & 100\% & 300, 10 \\
\hline
\end{tabular}
\end{table}

\subsection{Performance of the Account Allocation Algorithm}
After adjusting the shards to different TPS values, we measured the algorithm's time consumption and allocation performance. Fig. \ref{t-1} and \ref{t-3} show the maximum shard processing time and the cross-shard transaction ratio. P-Louvain consistently outperforms the others, with its advantage growing as $N_{TX}$ increases. For $N_{TX}$ = 30 million and $K$ = 8, P-Louvain reduces the cross-shard transaction ratio by 7.5
\% compared to TxAllo. Fig. \ref{t-2} and \ref{t-4} depict the time consumption of each algorithm and their allocation components. CLPA shows significant growth in execution time, while TxAllo and P-Louvain remain relatively stable. For $N_{TX}$ = 30 million and $K$ = 8, P-Louvain reduces the allocation execution time by 86\% compared to TxAllo. Thus, P-Louvain is both more time-efficient and more effective.

\begin{figure*}[t]
    \centering
    \begin{subfigure}[b]{0.26\textwidth}
        \centering
        \includegraphics[width=\textwidth,height=\dimexpr0.66\textwidth\relax]{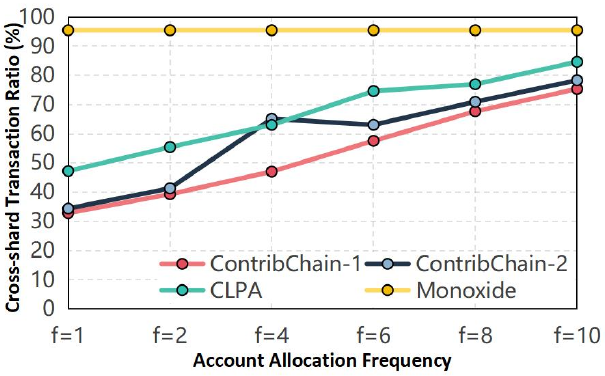} 
        \caption{Cross-shard TX Ratio}
        \label{crossratio-1}
    \end{subfigure}
    \hfill
    \begin{subfigure}[b]{0.23\textwidth}
        \centering
        \includegraphics[width=\textwidth]{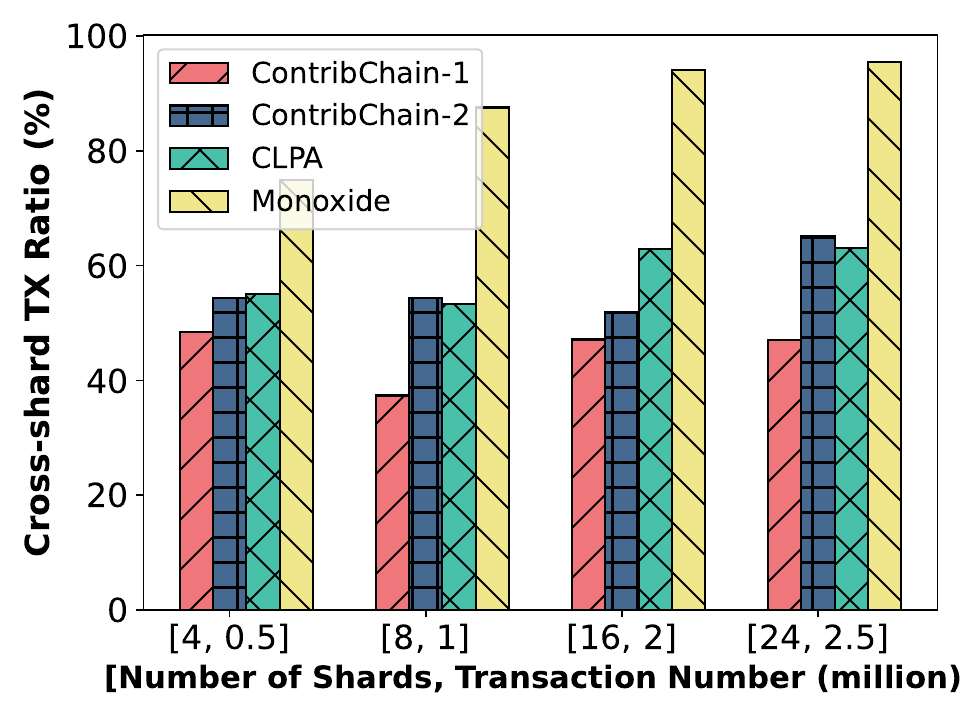} 
        \caption{Cross-shard TX Ratio}
        \label{crossratio-2}
    \end{subfigure}
    \hfill
    \begin{subfigure}[b]{0.24\textwidth}
        \centering
        \includegraphics[width=\textwidth]{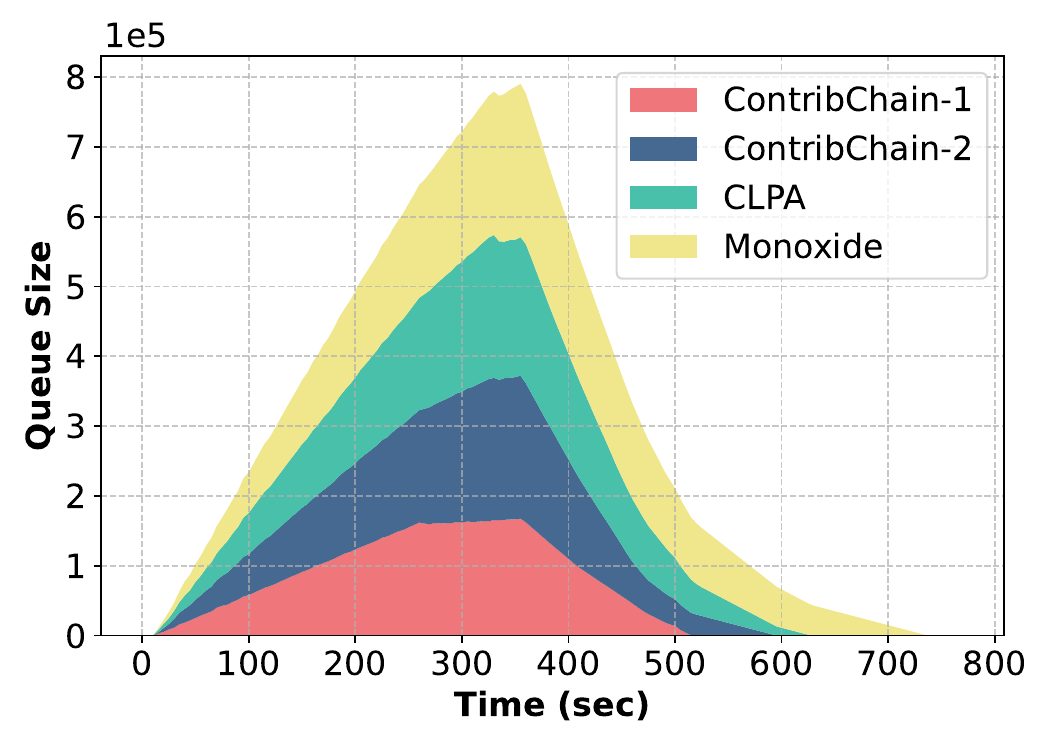} 
        \caption{TX arrival rate=1500 TXs/Sec}
        \label{queue_size-1}
    \end{subfigure}
    \hfill
    \begin{subfigure}[b]{0.24\textwidth}
        \centering
        \includegraphics[width=\textwidth]{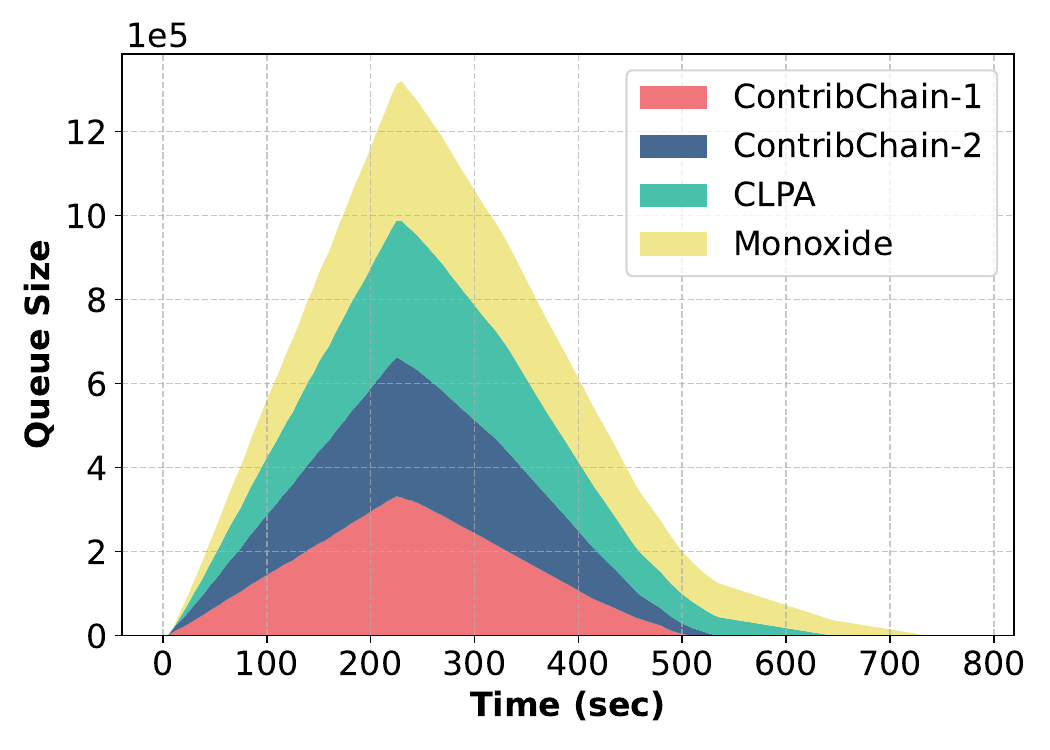} 
        \caption{TX arrival rate=2500 TXs/Sec}
        \label{queue_size-2}
    \end{subfigure}
    \caption{Cross-shard Ratio and queue size of the TX pool. In (a), $T_{NA}$=80 seconds. In (b), $N_{TX}$ = 500,000 and $K$=4.}
    \label{b}
\end{figure*}
\begin{figure*}[t]
    \centering
    \begin{subfigure}[b]{0.235\textwidth}
        \centering
        \includegraphics[width=\textwidth]{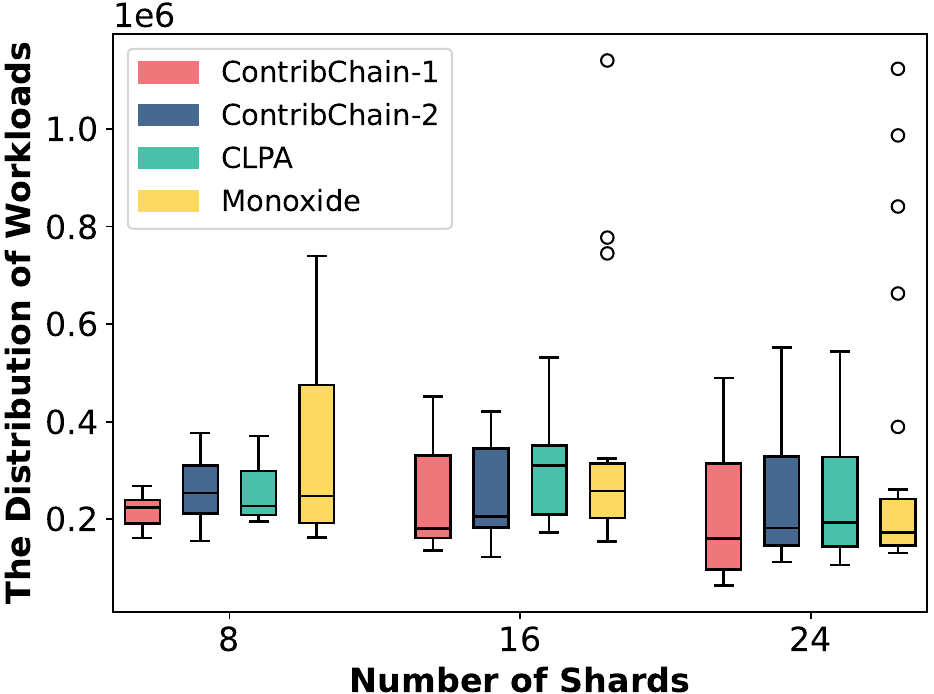} 
        \caption{Distribution of workloads}
        \label{load-tps-1}
    \end{subfigure}
    \hfill
    \begin{subfigure}[b]{0.280\textwidth}
        \centering
        \includegraphics[width=\textwidth]{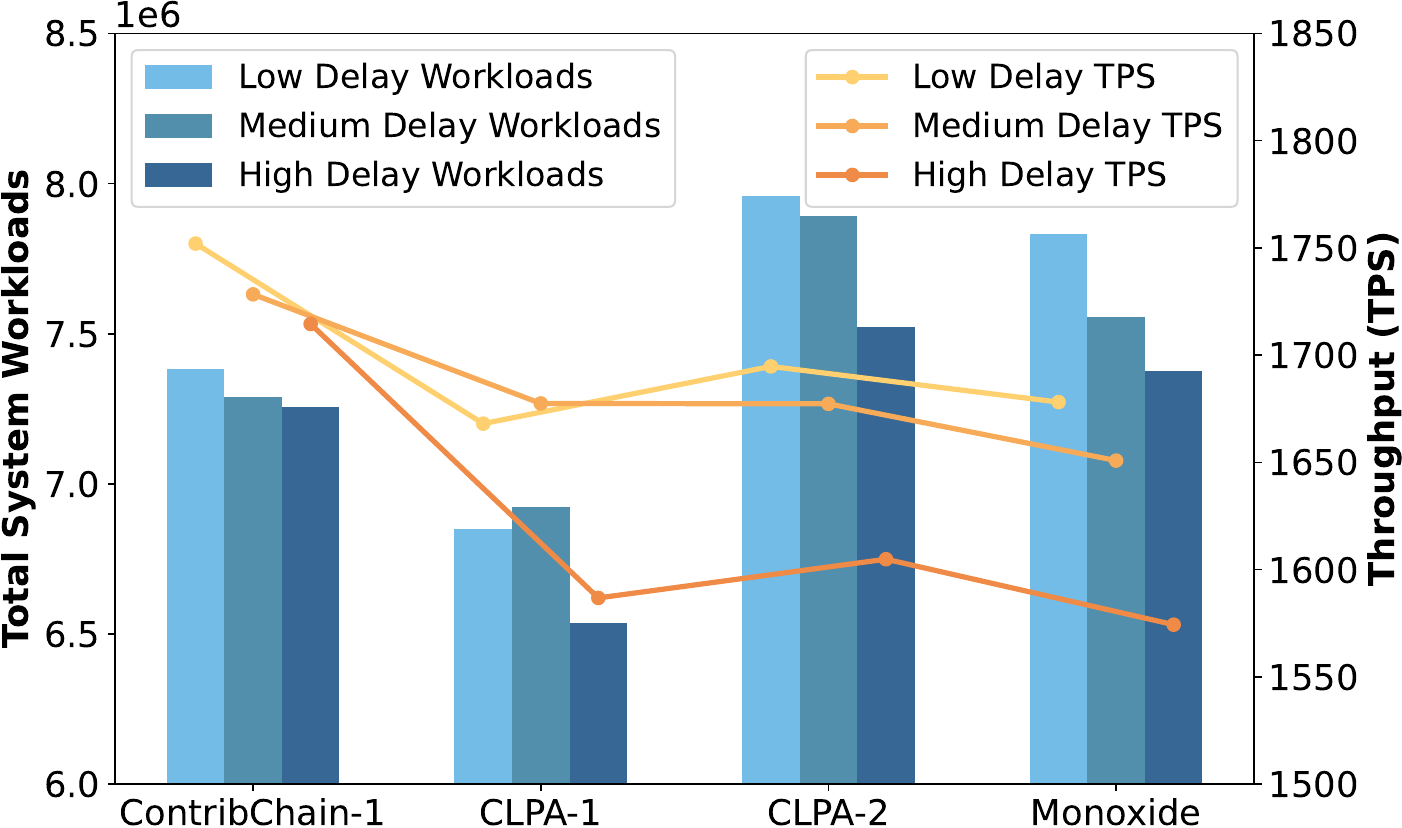} 
        \caption{Total system workloads}
        \label{load-tps-2}
    \end{subfigure}
    \hfill
    \begin{subfigure}[b]{0.227\textwidth}
        \centering
        \includegraphics[width=\textwidth]{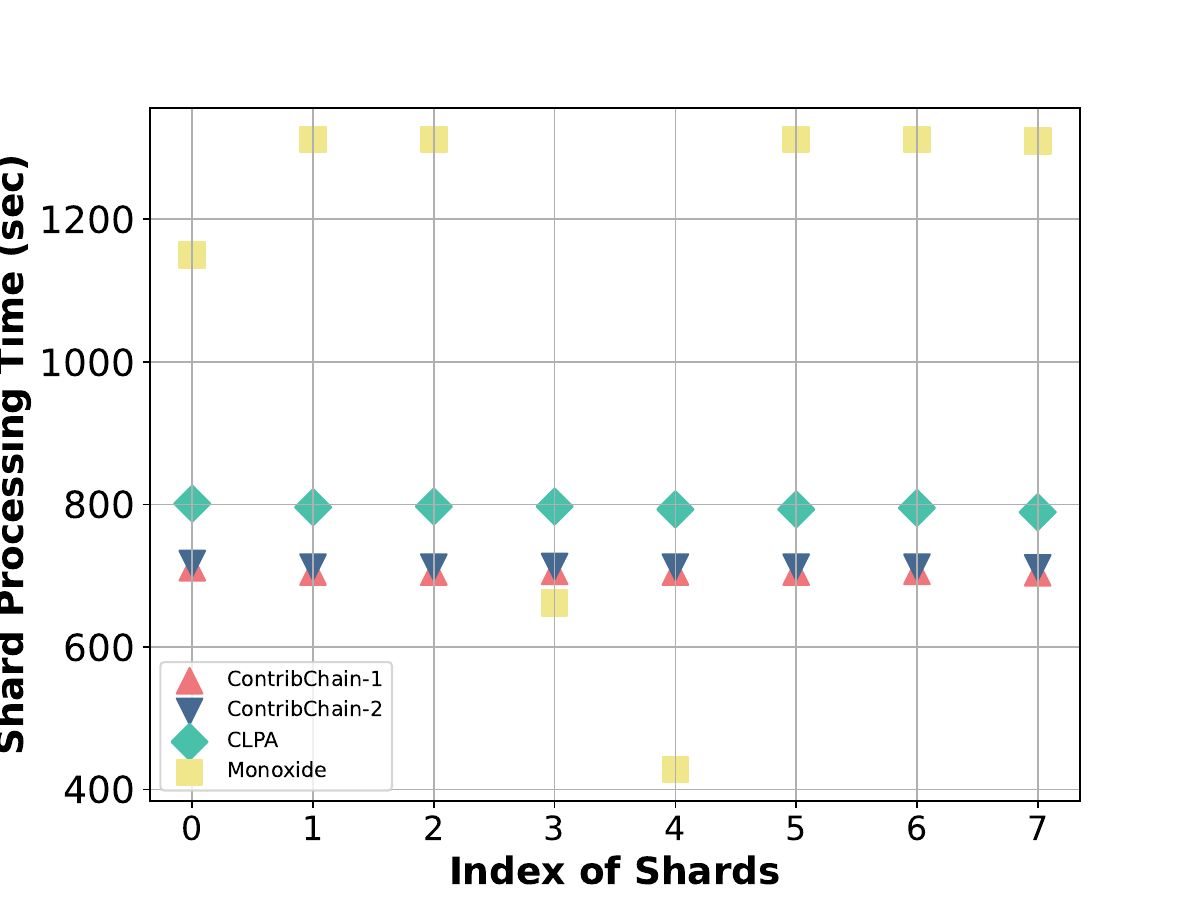} 
        \caption{Shard processing time}
        \label{10-a}
    \end{subfigure}
    \hfill
     \begin{subfigure}[b]{0.227\textwidth}
        \centering
        \includegraphics[width=\textwidth]{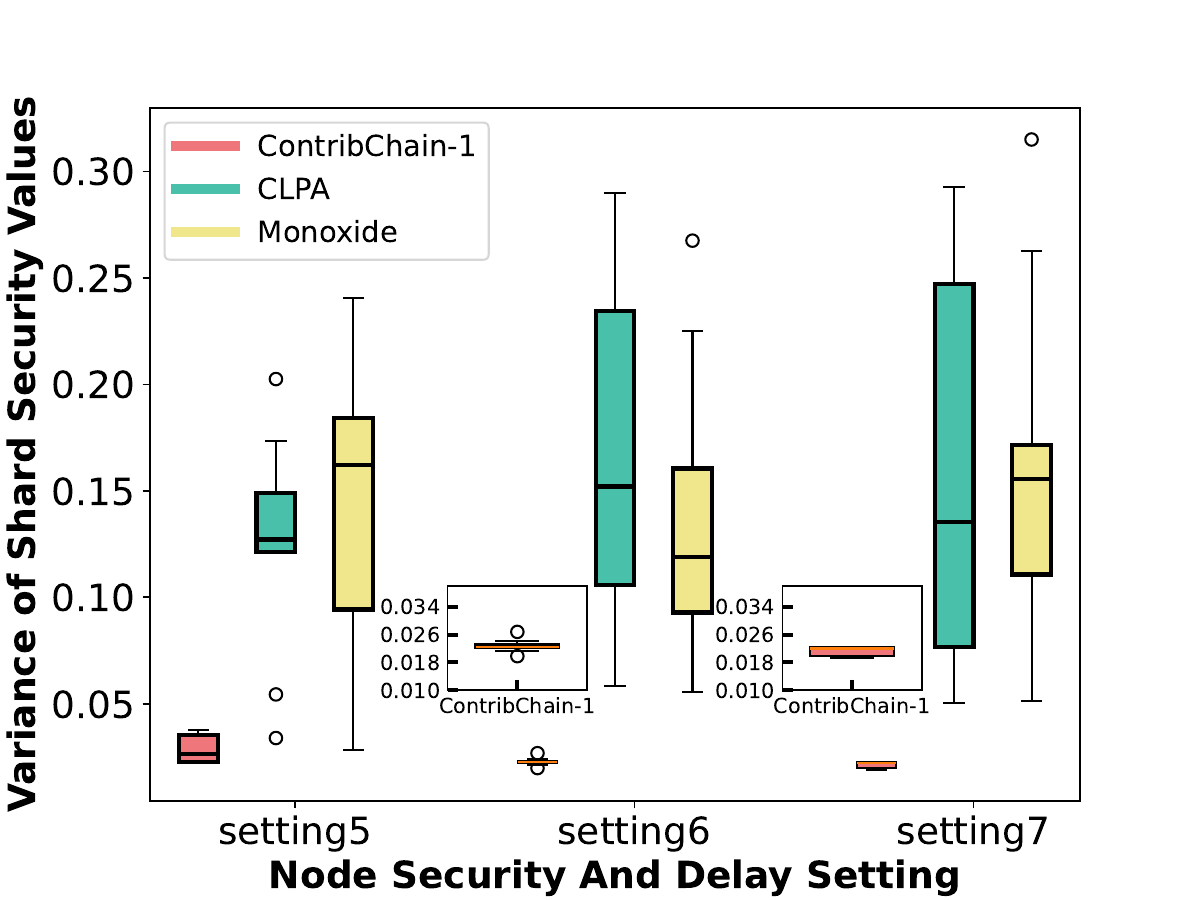} 
        \caption{Difference of shard security}
        \label{10-b}
    \end{subfigure}
    \caption{(a) and (b) show the impact of $K$ and delay settings on workload distribution and total system workload. (c) shows the processing time of shards. (d) shows the variance in shard security values under three node security and delay settings.}
    \label{fig:load-tps}
\end{figure*}

\subsection{System Throughput and TX Confirmation Latency}
We first set the account allocation frequency to every four epochs ($f$=4) and the node allocation period to 80 seconds ($T_{NA}$=80s). As shown in Fig. \ref{a-tps} and \ref{a-latency}, ContribChain-1 outperforms CLPA and Monoxide in terms of TPS and TX confirmation latency as both the TX arrival rate and $K$ increase. At a rate of 2000 tx/s and $K$=8, ContribChain-1 achieves a 92\% and 35.8\% improvement in TPS over Monoxide and CLPA, respectively. When the TX arrival rate reaches 3000 tx/s and $K$ increases from 16 to 24, the TPS of CLPA drops significantly, while ContribChain-1 exhibits minimal variation, demonstrating its superior scalability. Comparing ContribChain-1 and ContribChain-2, the former shows better performance, highlighting the synergistic enhancement provided by P-Louvain and NACV cooperatively.

Next, we adjust $f$ while $T_{NA}$=80s. As seen in Fig. \ref{b-tps} and \ref{b-latency}, Monoxide, without account allocation, serves as the yellow baseline. Overall, the TPS increases as $f$ rises from 1 to 4. However, as $f$ increases beyond 4, the TPS begins to decrease due to more frequent account allocation, which negatively impacts consensus time, while a larger $f$ increases the cross-shard transaction ratio. Consequently, we set the default value of $f$ to 4.

We also evaluated the system performance under different node delays, running 8 epochs of 200 seconds each, with a TX arrival rate of 2500 tx/s. For CLPA-1, $f$ is set to 1, while for CLPA-2 and ContribChain-1, $f$ is set to 4. As shown in Fig. \ref{C-TPS} and \ref{C-Latency}, ContribChain-1 exhibits minimal fluctuation in TPS and TX confirmation latency as delays increase. Despite having higher TX confirmation latency than CLPA-1 at low delays, ContribChain-1 maintains the highest TPS, benefiting from fewer account migrations and reduced overhead. 

Finally, we adjust both the node security and delay settings (Table \ref{settings}). The delay of node $i$ in shard $j$ is $L_s*j+L_n*i$. With $N_{TX}$ = 1 million, a TX arrival rate of 2000 tx/s, and $T_{NA}$ = 80 seconds, Fig. \ref{D-TPS} and \ref{D-Latency} show that, compared to CLPA and Monoxide, ContribChain-1 and ContribChain-2 maintain more stable TPS and TX confirmation latency, performing better even as node security and performance degrade.

\subsection{Cross-shard TX Ratio and TX Pool Queue Size}
\textbf{Cross-Shard TX Ratio:} We evaluate the impact of $f$, $K$, and $N_{TX}$ on the cross-shard TX ratio. As shown in Fig.  \ref{crossratio-1} and \ref{crossratio-2}, ContribChain-1 consistently exhibits the lowest cross-shard TX ratio. At $f$=6, ContribChain-1 reduces the cross-shard TX ratio by 16\% compared to CLPA. Moreover, as $K$ and $N_{TX}$ increase, the cross-shard TX ratio in ContribChain-1 remains nearly constant, indicating strong scalability.

\textbf{TX Pool Queue Size:} We also analyze the TX pool queue size for $N_{TX}$ = 500,000 and $K$=4. Transactions are injected into the TX pool at rates of 1500 tx/s and 2500 tx/s until all transactions are processed. Fig. \ref{queue_size-1} and \ref{queue_size-2} show that once the TXs are fully injected, the queue size decreases. Among all methods, ContribChain-1 has the shortest queue, with the lowest peak and the fastest time to empty the queue, demonstrating superior efficiency in handling transaction backlogs.

\subsection{Balance among Shards}
\textbf{Workload Balance:}
We assess the workload distribution among shards for $K$=8, 16, and 24. Fig. \ref{load-tps-1} reveals that Monoxide suffers from poor workload distribution, exhibiting many outliers. In contrast, ContribChain-1 shows a more uniform load distribution with no outliers, indicating better workload balance. We then evaluate total workload and TPS after running 8 epochs with different node delay settings (Settings 2, 3, and 4 in Table \ref{settings}). The parameters are 8 shards, 2500 tx/s, and 200 seconds per epoch. Fig. \ref{load-tps-2} shows that CLPA-1 achieves lower TPS and higher workload due to frequent account allocations, which reduce consensus time. Under the same $f$, ContribChain-1 achieves lower workload than CLPA-2, owing to fewer cross-shard transactions. As node delays increase, ContribChain-1 demonstrates superior self-regulation compared to other algorithms.

\textbf{Stress and Security Balance:}
With $N_{TX}$ fixed, we analyze the processing time of all shards (Fig. \ref{10-a}). Through inter-shard comparison, we can observe a significant disparity in processing times in Monoxide. The transaction backlog problem depicted in Fig. \ref{1-Problems} occurs in shards other than shard 3, 4. While ContribChain-1 have more uniform and shorter processing time, showing better stress balancing capabilities. Fig. \ref{10-b} shows the distribution of the variance of shard security values across different epochs. The shard security value is calculated from the average of the security contribution values of the nodes within the shard. We study this distribution in three node security and delay settings (Setting 5, 6, 7 in Table \ref{settings}). ContribChain-1 consistently maintains low and stable variance after each node allocation, indicating excellent security balancing capabilities. In contrast, CLPA and Monoxide exhibit larger and more fluctuating variances.

\section{Conclusion}
ContribChain introduces a stress-balanced sharding protocol for blockchain systems that dynamically evaluates node performance and security through the update of node contribution values. Additionally, we propose novel account and node allocation algorithms to achieve optimal stress balance across the network. Experimental results demonstrate that ContribChain outperforms existing protocols in key metrics, including transaction throughput, cross-shard transaction ratio, confirmation latency, TX pool queue size, and stress balance. In future work, we aim to further enhance the system's adaptability by incorporating dynamic adjustments to the number of shards and the frequency of account allocation.

\section*{Acknowledgment}
This work was partially supported by the National Key R\&D Program of China (2021YFB2700300), the National Natural Science Foundation of China (62141605, 62372493), the Beijing Natural Science Foundation (Z230001), the Beijing Advanced Innovation Center for Future Blockchain and Privacy Computing (GJJ-23-001, GJJ-23-002),  the China Postdoctoral Fellowship Fund 373500, the Beihang Dare to Take Action Plan KG16336101, and the science and technology project of State Grid Corporation of China (5400-202255416A-2-0-ZN).






%


\bibliographystyle{IEEEtran}
\bibliography{references} 
\end{document}